\documentclass[twocolumn,pra,superscriptaddress,eqsecnum]{revtex4}
\usepackage{amsmath}
\usepackage{amsfonts}
\usepackage{amssymb}
\usepackage{amstext}
\usepackage[sort&compress]{natbib}
\usepackage{graphicx}           
\usepackage{amsmath}            
\usepackage{wasysym}
\usepackage{verbatim}           
\usepackage{amsfonts,amssymb,latexsym,amscd}
\usepackage{color}
\usepackage{times}              
\usepackage{amsthm}

\theoremstyle{plain}
\newtheorem{theorem}{Theorem}
\newtheorem{definition}{Definition}

\newtheorem{proposition}[theorem]{Proposition}

\newtheorem*{remark}{Remark}
\newtheorem{example}{Example}

\begin{document}

\title{Graph States for Quantum Secret Sharing}

\author{Damian Markham}
\email{djhm@pps.jussieu.fr}
\affiliation{Universit\'{e} 
    Denis Diderot, 175 Rue du Chevaleret, 75013 Paris, France}
\affiliation{Department of Physics,
Graduate School of Science,
  University of Tokyo, Tokyo 113-0033, Japan}
\author{Barry C. Sanders}
\affiliation{Institute for Quantum Information Science, University of Calgary, Alberta T2N 1N4,
Canada}



\begin{abstract}
We consider three broad classes of quantum secret sharing with and without eavesdropping
and show how a graph state formalism unifies otherwise disparate quantum secret sharing
models. In addition to the elegant unification provided by graph states, our approach provides
a generalization of threshold classical secret sharing via insecure quantum channels beyond
the current requirement of 100\% collaboration by players to just a simple majority in the case of
five players. Another innovation here is the introduction of embedded protocols within
a larger graph state that serves as a one-way quantum information processing system.

\end{abstract}

\maketitle
\section{Introduction}
\label{sec:introduction}

Secret sharing is one of the most important information-theoretically secure cryptographic protocols
and is germane to online auctions, electronic voting, shared electronic banking,
and cooperative activation of bombs.  Although first formulated
and solved in classical information terms~\cite{Sha79}, two directions have been followed
in the quantum case: classical secret sharing with quantum enhancement of channels to protect
against eavesdropping~\cite{HBB99} (and experimentally realized~\cite{GRTZ02})
and protecting quantum information as an application of quantum error correction~\cite{CGL99}
(and experimentally realized in the `continuous variable' Gaussian state case~\cite{TS02}).

In secret sharing, a dealer wishes to transmit a secret, which can be a bit string or a qubit string
and deals an encoded version of this secret to $n$ players such that some subsets
of players can collaborate to reconstruct the secret and all
other subsets are denied any information whatsoever
about the secret.
The access structure is the set of all subsets of players that can obtain the secret,
and the adversary structure is the set of all subsets of players that are completely denied the secret.

In threshold secret sharing, each player receives precisely one equal
share of the encoded secret, and a threshold number of any players~$k$ can collaborate
to reconstruct the secret whereas every subset of fewer than~$k$ players is denied any
information at all: this scheme is referred to as $(k,n)$ threshold secret sharing and is
a primitive protocol by which any information-theoretic secret sharing protocol can be
achieved.

In each case the channel can be classical or quantum depending on context.
A private channel is an authenticated channel that is impervious to eavesdropping.
A public channel is an authenticated channel that is open to eavesdropping.
Mathematically channels are described by completely positive trace-preserving mappings,
even for classical channels as classical information processing can be embedded into
a quantum framework.

Our goals here are
\begin{itemize}
	\item [(i)]
	to formulate a secret sharing problem that allows both classical
	and quantum channels, which can be private or public,
	\item [(ii)]
	to introduce three subproblems that
	show how the original classical secret sharing and the two versions of quantum secret sharing
	fit into the general problem and highlight the differences between these subproblems,
	\item [(iii)]
	develop a graph state formalism~\cite{HDERNB06}
	that unites these subproblems into one elegant framework,
	\item [(iv)]
	extend one class of subproblems concerning classical secret sharing in insecure
	public channels beyond the existing security proofs that require 100\% collaboration
	by the players, and
	\item [(v)]
	introduce embedded secret sharing protocols for graph states within one-way
	quantum computing, which could serve as a
	first step towards implementing integrated secret sharing protocols within quantum computing,
	e.g. distributed measurement-based quantum computation (MBQC)~\cite{RB01}.
\end{itemize}

\section{Secret sharing problem and subproblems}
\label{sec:problem}

To begin we formulate the general problem. The basic unit of classical information is
the bit, corresponding to~$\{0,1\}$, and the basic unit of quantum information is the
qubit, corresponding to ~$\mathcal{H}_2=\text{span}\{|0\rangle,|1\rangle\}$. Any
message can be encoded in a finite string of bits if the information is classical or
in a finite string of qubits (perhaps entangled and could be pure or mixed) if the
`information' is quantum.
\begin{quote}
    \textbf{Secret Sharing Problem:} A dealer holds a secret~$S$, which is either
    a bit or a qubit,
    and must send this secret to $n$ players such that any $k$ or more players
    can reconstruct the secret, and all sets of fewer than~$k$ players as well
    as eavesdroppers are denied any access whatsoever to the secret.
\end{quote}
\begin{remark}
	A secret that is longer than one bit or one qubit can be shared by a distributed protocol
	that shares the string one bit or qubit at a time.
\end{remark}
Our new and general formulation of the secret sharing problem is meant to be helpful in that
it incorporates various scenarios of secret sharing with quantum or classical
channels that can be public or private, and the three major secret sharing
areas of study~\cite{Sha79,HBB99,CGL99} are subproblems of this overarching problem.
\begin{enumerate}
\item [{\sf CC}]
	\underline{C}lassical secret sharing with private
	channels between the dealer and each player
	and private \underline{C}lassical channels shared between each pair of players;
\item [{\sf CQ}]
	\underline{C}lassical secret sharing with public channels between
	the dealer and each player and either \underline{Q}uantum or classical channels
	shared between each pair of players
\item [{\sf QQ}]
	\underline{Q}uantum secret sharing wherein the dealer shares quantum channels with each player, and these
	channels can be private or public, and the players share either \underline{Q}uantum or classical
	private channels between each other.
\end{enumerate}

As secret sharing is information-theoretically secure, the distribution of a classical secret message
is completely secure without the need for quantum information, in contrast to quantum key
distribution~\cite{BB84}, which catapults key distribution over public channels to information-theoretic
security from the classical limited computational-security guarantee.
Thus case~{\sf CC}~\cite{Sha79} is resolved fully by classical information processing,
but we embed {\sf CC} into a graph state description.

Case~{\sf CQ} considers quantum-enhanced protection from eavesdropping over public
channels and was first studied for $(2,2)$ threshold secret sharing~\cite{HBB99,KKI99,Chen04}
and extended to $(n,n)$ threshold secret sharing~\cite{Xiao04}.
The quantum enhancement is achieved by random key distribution
enabled via quantum purification.

The third case~{\sf QQ} deals with the sharing of quantum information~\cite{CGL99},
as opposed to classical information in the first two subproblems identified above.
To distinguish this case from the first two cases, the term `quantum state sharing'
is sometimes used~\cite{TS02}. In contrast to~{\sf CQ}, which is only solved
for the~$(n,n)$ threshold secret sharing case, this quantum information version in
private channels is fully general in the sense that the general $(k,n)$ case has
been solved as an application of quantum error correction.

In summary our graph state formalism unifies known results concerning
$(n,n)$~{\sf CQ} plus~{\sf QQ} for higher-dimensional stabilizer states, and
adds new results, including $(3,5)$~{\sf CQ}.
We expect that our new approach to all forms of quantum secret sharing in
the literature will enable further advances because of the elegance and
unifying properties of this formalism.

\section{Graph States}
\label{sec: graphstates}

Graph states provide a superb resource for secret sharing, as we show. The advantages of using graph states in this work are threefold. First, they are the most readily available multipartite resource states in the laboratory at present, with graph states of up to 6 qubits already built and used for information processing experimentally~\cite{WRRSWVAZ05,LGZGZYGYP05}, and many proposals for their implementation in various systems. Second, they are natural candidates for integration of different tasks, since they are the main state resource in almost all multipartite quantum information processing tasks, including measurement-based quantum computation and error correction (hence the large effort to make them). Third, they allow for an elegant graphical representation which offers an intuitive picture of information flow, and also allow the possible use of graph theoretic results to aid proofs and understanding.

In this work we employ graph states in the standard form~\cite{HDERNB06}
plus two extensions. These extensions are actually quite simple (modification via local unitaries) and are intended to allow intuitive graphical understanding of how information is encoded and spread over the graphs.
The idea is that the secret will be encoded onto classical labels placed on vertices of the graph representing local operations. The inherent entanglement of the graph states allows these labels to be shifted around, allowing us to see graphically which sets of players can access the secrets and which cannot (explicitly written as properties P1-P4, see e.g. Fig.~\ref{fig:GHZMexample}). We will give several examples along the way to make it clear.

We will begin by defining these extended `labeled' graph states in Subsec.~\ref{subsec:labeled graph states}, followed by describing the graphical rules of how they are used to spread information and how it can be accessed in Subsec.~\ref{subsec:properties}, finishing with how they change under measurements in Subsec.~\ref{subsec:measurements}.

\subsection{Labeled graph states}
\label{subsec:labeled graph states}

A pure graph state is a state in the Hilbert space~$\mathcal{H}_2^{\otimes n}$,
i.e.\ a string of~$n$ qubits. For $|\pm\rangle=(|0\rangle\pm |1\rangle)/\sqrt{2}$
and an undirected graph $\mathsf{G}=(\mathsf{V},\mathsf{E})$ comprising $n$ vertices, with
\begin{equation}
	\mathsf{V}=\{\mathsf{v}_i\},\
	\mathsf{E}=\{\mathsf{e}_{ij}=(\mathsf{v}_i,\mathsf{v}_{j})\}
\end{equation}
the set of $n$~vertices and the set of edges, respectively.
\begin{definition}
Two players~$i$ and~$j$ are `neighbours' iff there exists an edge~$\mathsf{e}_{ij}$
that connects their respective vertices~$\mathsf{v}_i$ and~$\mathsf{v}_j$.
\end{definition}
\noindent The set of $i$'s neighbours is denoted~$N_i$.

A graph state is created from an initial state
\begin{equation}
\label{eq:+n}
	\left|+\right\rangle^{\otimes n}
		=H^{\otimes n}\left|0\right\rangle^{\otimes n}, \,
	H=\left|+\right\rangle\left\langle 0\right|
		+\left|-\right\rangle\left\langle 1\right|,
\end{equation}
with~$H$ the Hadamard transformation,
by applying the two-qubit controlled-phase gate
\begin{equation}
\label{eq:CZ}
	\text{CZ}|\epsilon\epsilon'\rangle=(-1)^{\epsilon\epsilon'}|\epsilon\epsilon'\rangle,\,
	\epsilon,\epsilon'\in\{0,1\},\,|\epsilon\epsilon'\rangle\in\mathcal{H}_2^{\otimes 2}
\end{equation}
to all pairs of qubits whose corresponding vertices on the graph
are joined by an edge and not to any pair not connected by an edge.
The commutativity of all CZ operations implies that the
order of applying CZ gates is unimportant:
\begin{equation}
\label{eq:graphstate}
	|\mathsf{G}\rangle
		= \prod_{\mathsf{e}\in\mathsf{E}} \text{CZ}_\mathsf{e}|+\rangle^{\otimes n}.
\end{equation}

We now modify the graph labeling. The usual graph state has each node labeled by
its vertex index; in other words the label of vertex~$\mathsf{v}_i$ is the index~$i$, which
has a bit string length of $O(\log n$). We append to this label three more bits so the
label of vertex~$\mathsf{v}_i$ is $(i,\ell_{i1},\ell_{i2},\ell_{i3})$ for each $\ell_{ij}\in\{0,1\}$.

The first two labels are used to describe classical information spread over the state.
Over all vertices, we define the vectors
$\vec{\ell}_{i\star}=(\ell_{i1},\ell_{i2})$ for the~$i^\text{th}$ vertex,
$\vec{\ell}_{\star j}=(\ell_{1j},\ell_{2j},\ldots,\ell_{nj})$ for the~$j^\text{th}$ bit over all~$n$ vertices,
and
\begin{equation}
\label{eq:vecell}
	\vec{\ell}=(\vec{\ell}_{1\star},\vec{\ell}_{2\star},\ldots,\vec{\ell}_{n\star}).
\end{equation}
Graphically these labels will be attached to their vertices and used to represent the spreading of classical information. If a vertex $\mathsf{v}_i$ has no label, this is equivalent to setting $\vec{\ell}_{i\star}=(0,0)$.

The third additional bit $\ell_{i3}$ occurs in the security of protocols, and is not relevant to the classical information transmitted. For the ease of graphical manipulation of the encoded information, we absorb this bit into the graph itself. Graphically the third bit~$\ell_{i3}$ is depicted as the vertex~$\mathsf{v}_i$
being either a $\ocircle$ (the set of which is denoted $\mathsf{V}^\ocircle$), for $\ell_{i3}=0$, or a $\square$ (the set of which is denoted $\mathsf{V}^\square$),
for $\ell_{i3}=1$.
The vertex set is thus a union of two types of vertices,
\begin{equation}
	\mathsf{V}=\mathsf{V}^\ocircle \cup \mathsf{V}^\square.
\end{equation}
We thus have the extended graphs $\mathcal{G}=(\mathsf{V}^\ocircle, \mathsf{V}^\square, \mathsf{E})$.

These extra labels imply the action of local unitary operations (hence maintaining entanglement properties) via the following definition.
\begin{definition}
\label{def:labeledgraphstate}
	The `labeled graph state',
	is obtained from the graph state~$|G\rangle$ by
	\begin{align}
	\label{eq:labeledgraphstate}
		\left|\mathcal{G}_{\vec{\ell}}\right\rangle
			=& \bigotimes_i \left( X_i^{\ell_{i1}} Z_i^{\ell_{i2}} \right) \left|\mathcal{G}\right\rangle \nonumber \\
        \left|\mathcal{G}\right\rangle
            =& \bigotimes_{j|\mathsf{v}_j\in V^\square}S_j \left|\mathsf{G}\right\rangle
    \end{align}
	for
	\begin{align}
		X=&|0\rangle\langle 1|+|1\rangle\langle 0|,	\nonumber \\
		Z=&|0\rangle\langle 0|-|1\rangle\langle 1|,	\\
		S=&|0\rangle\langle 0|-i|1\rangle\langle 1|.	\nonumber
	\end{align}
\end{definition}
\begin{remark}
The partial phase shift gate~$S$ is used to perform an $X \leftrightarrow Y$
basis transformation
for protection against eavesdropping, analogous to the BB84 strategy~\cite{BB84}.
\end{remark}

\bigskip

For encoding purposes, only local $Z$ gates are required, so we introduce another
kind of graph state.
\begin{definition}
	The `encoded graph state' is
	\begin{equation}
	\label{eq:InformationalState}
		|\mathcal{G}_{\vec{\ell}_{\star 2}}\rangle
			= \bigotimes_i Z_i^{\ell_{i2}}|\mathcal{G}\rangle.
	\end{equation}
\end{definition}
Thus the encoded graph state is also a labeled graph state with $\ell_{1i}=0$ for all $i$, $|\mathcal{G}_{\vec{\ell}_{\star 2}}\rangle~=~|\mathcal{G}_{\vec{\ell}
	=(0,\ell_{12}, 0,\ell_{22},\ldots ,0,\ell_{n2}) }\rangle$.
Encoded graph states exhibit the orthonormality property
\begin{equation}
\label{eq:Gorthonormality}
	\left\langle\mathcal{G}_{\vec{\ell}_{\star 2}}\big|\mathcal{G}_{\vec{\ell}'_{\star 2}}\right\rangle
		=\delta_{\vec{\ell}_{\star 2}\vec{\ell}'_{\star 2}}.
\end{equation}


\bigskip

Encoded graph states~$|\mathcal{G}_{\vec{\ell}_{\star 2}}\rangle$
can be elegantly expressed in the stabilizer formalism.
For $Y=\text{i}XZ$ and stabilizer operators for any $i^\text{th}$ vertex denoted by
\begin{equation}
\label{eq:eigenoperators}
	K_i^{\ocircle} = X_i\otimes_{\mathsf{e}_{i,j}\in\mathsf{E}}Z_j,\,
	K_i^{\square} = Y_i\otimes_{\mathsf{e}_{i,j}\in\mathsf{E}}Z_j,
\end{equation}
(for circular and square vertices respectively) the encoded state $|\mathcal{G}_{\vec{\ell}_{\star 2}}\rangle$ is completely specified by eigenequations
\begin{equation}
\label{eq:eigenequations}
	K_i^{\ocircle,\square} |\mathcal{G}_{\vec{\ell}_{\star 2}}\rangle
		= (-1)^{\ell_{i2}} |\mathcal{G}_{\vec{\ell}_{\star 2}}\rangle\,
			\forall i \in \mathsf{V}.
\end{equation}

\begin{figure}
	\includegraphics[width=3.5cm]{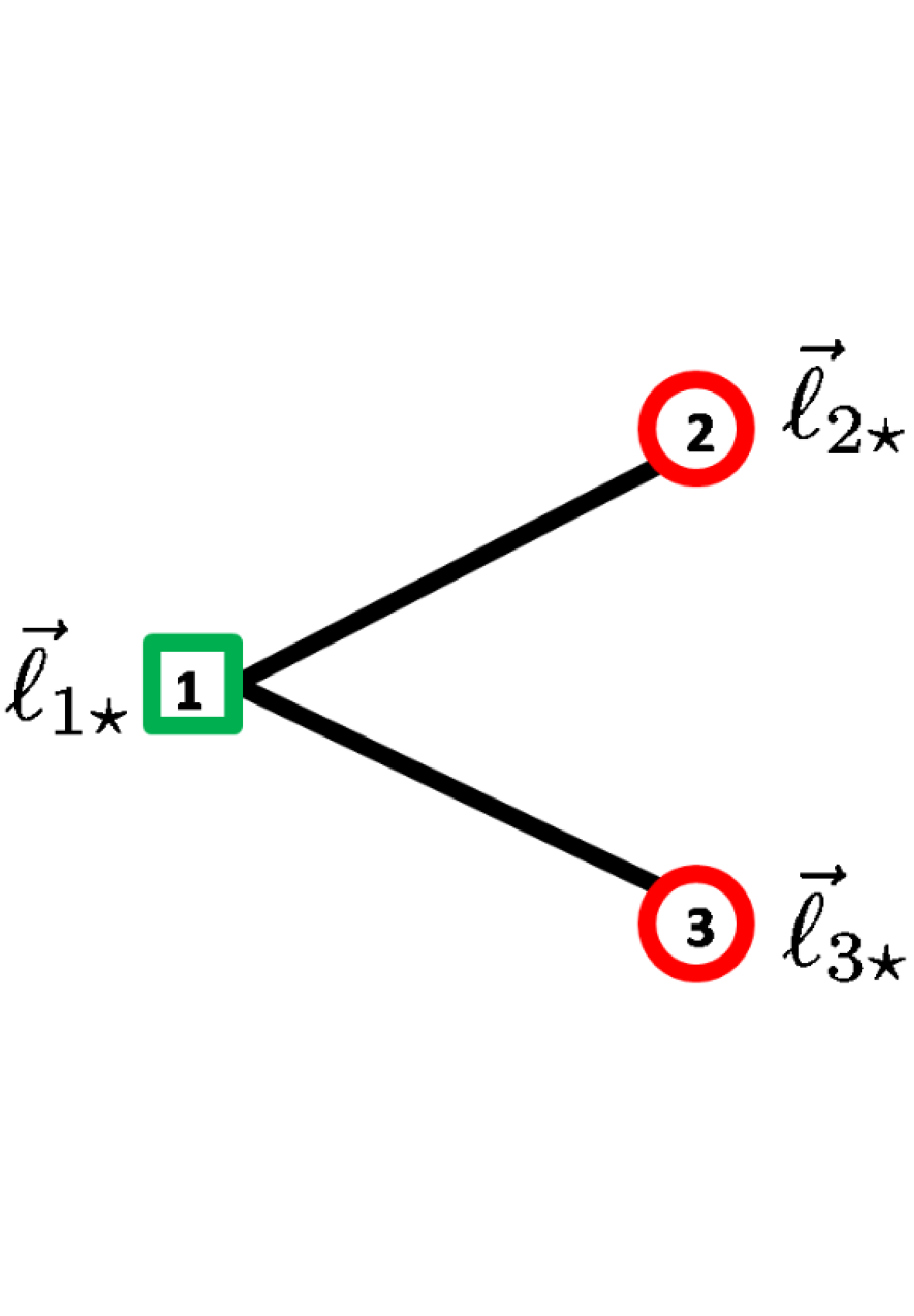}
	\caption{\label{fig:GHZConj}
	(Color
 online) Labeled graph with vertex~$\mathsf{v}_1$,
	depicted as~1 inside a~$\square$, which indicates that~$S$ has been
	applied to this vertex, and vertices~$\mathsf{v}_2$ and~$\mathsf{v}_3$,
	depicted as the numerals~2 and~3 inside~$\ocircle$.
		The labels $\vec{\ell}_{i\star}=(\ell_{i1},\ell_{i2})$ indicate that operation $X^{\ell_{i1}}Z^{\ell_{i2}}$ have
		been applied to each qubit~$j$, respectively.
		For $\vec{\ell}_{i\star}\equiv (0,\ell_{i2})\,\forall i$, this labeled graph state
		corresponds to the encoded state $|\mathcal{G}_{\vec{\ell}_{\star2}}\rangle~=~Z_1^{\ell_{12}}\otimes~Z_2^{\ell_{22}}\otimes~Z_3^{\ell_{32}}\left(|0++\rangle+i|1--\rangle\right)/\sqrt{2}$ (Eqn.~(\ref{eq:3EG})).
	}
\end{figure}
The following two examples, for three and for four qubits, respectively,
explain how the stabilizers define the encoded graph states.

\begin{example}
The three-qubit labeled graph state presented in Fig.~\ref{fig:GHZConj}
is the encoded graph state
\begin{align} \label{eq:3EG}
|G_{\vec{\ell}_{\star 2}}\rangle = Z_1^{\ell_{12}} \otimes Z_2^{\ell_{22}} \otimes Z_3^{\ell_{32}}
\left(\frac{|0++\rangle +i|1--\rangle}{\sqrt{2}}\right),
\end{align}
and is the unique common eigenstate of
\begin{align}
	K_1^\square=&Y_1 \otimes Z_2 \otimes Z_3 \nonumber \\
	K_2^\ocircle =&Z_1 \otimes X_2 \otimes \openone_3 \nonumber \\
	K_3^\ocircle =&Z_1 \otimes \openone_2 \otimes X_3,
\end{align}
with eigenvalues $\vec{\ell}_{\star 2}=(l_{12},l_{22},l_{32})$.
\end{example}

A four-qubit encoded graph state is given in the following example.
First we define the~$n$-qubit Greenberger-Horne-Zeilinger-Mermin ($n$GHZM)
graph and corresponding $n$GHZM graph state.

\begin{definition}
	The \emph{$n$GHZM graph} for vertex~$\mathsf{v}_i$ is the degree~$n$ graph
	\begin{equation}
	\label{eq:nGHZMi}
		\mathcal{G}_{n\text{GHZM}}=\left(\mathsf{V}=\{\mathsf{v}_j\},
			\mathsf{E}=\{(\mathsf{v}_1,\mathsf{v}_{j\neq i})\}\right),
	\end{equation}
	and the \emph{$n$GHZM graph state} is the state corresponding to
	$|\mathcal{G}_{n\text{GHZM}}\rangle$.
\end{definition}

\begin{example}
\label{ex:4GHZM}
	The $4$GHZM graph is depicted in Fig.~\ref{fig:GHZMexample}(a),
	and the corresponding encoded $4$GHZM graph state is
	\begin{align}
	\label{eq:4GHZMdependencyEG}
		|\mathcal{G}_{\vec{\ell}_{\star 2}}\rangle
			=& Z^{\ell_{12}} \otimes Z^{\ell_{22}} \otimes Z^{\ell_{32}}
				\otimes Z^{\ell_{42}}|\mathcal{G}\rangle, \\
			|\mathcal{G}\rangle=&\frac{1}{\sqrt{2}}\left( |0+++\rangle + |1---\rangle \right),
					\nonumber
	\end{align}
and is fully defined by the four stabilizers
	\begin{align}
	\label{eq:4GHZMdependencyEG eigen-ops}
		K_1^\ocircle =&X \otimes Z \otimes Z \otimes Z,\,
		K_2^\ocircle =Z \otimes X \otimes \openone \otimes \openone, \nonumber \\
		K_3^\ocircle =&Z \otimes \openone \otimes X \otimes \openone,\,
		K_4^\ocircle =Z \otimes \openone \otimes \openone \otimes X,
	\end{align}
with eigenvalues $\vec{\ell}_{\star 2}=(l_{12},l_{22},l_{32},l_{42})$.
\end{example}

\subsection{Dependence and Access}
\label{subsec:properties}

We first consider sharing a classical secret;
sharing quantum secrets follows from this as we will see later.
Each qubit represented by a vertex in the graph
is held by one player, and the classical secret is encoded into the label~$\vec{\ell}_{\star 2}$.
A subset of players is represented by a vertex subset $\mathsf{V'}\subset\mathsf{V}$.

For general labeled graph states, all accessible information for $\mathsf{V}'$ resides in the reduced state obtained
from the labeled graph state~(\ref{eq:labeledgraphstate}):
\begin{equation}
\label{eq;reducedstate}
	\rho_{\mathsf{V'}\vec{\ell}}
		= \text{Tr}_{\mathsf{V/V'}}|\mathcal{G}_{\vec{\ell}}\rangle \langle\mathcal{G}_{\vec{\ell}}|.
\end{equation}

In the context of secret sharing, it is important to know what information about~$\vec{\ell}$
is contained in $\rho_{\mathsf{V'}\vec{\ell}}$. The information shared between $\rho_{\mathsf{V'}\vec{\ell}}$ and
$\vec{\ell}$ exhibits two important properties:
\begin{itemize}
\item [A1]
	\emph{Bit dependence:--}
	$\rho_{\mathsf{V'}\vec{\ell}}$ is \emph{dependent} on bit~$k$ if~$\rho_{\mathsf{V'}\vec{\ell}}$
	is not invariant under a flip of the~$k^\text{th}$ bit.
\item [A2]
	\emph{Bit access:--}
	Bit~$k$ is \emph{accessible} if there exists a measurement protocol
	such that one copy of $\rho_{\mathsf{V'}\vec{\ell}}$ reveals the value of bit~$k$
	with certainty.
\end{itemize}
A2$\Rightarrow$A1, but A1$\nRightarrow$A2.
Given dependency on a bit, it is especially useful to know what information can be obtained
by $\mathsf{V}'$ via local operations and classical communication (LOCC) or via quantum channels.

\begin{figure}
\scalebox{0.45}{\includegraphics*[0.5cm,8.5cm][21cm,21.5cm]{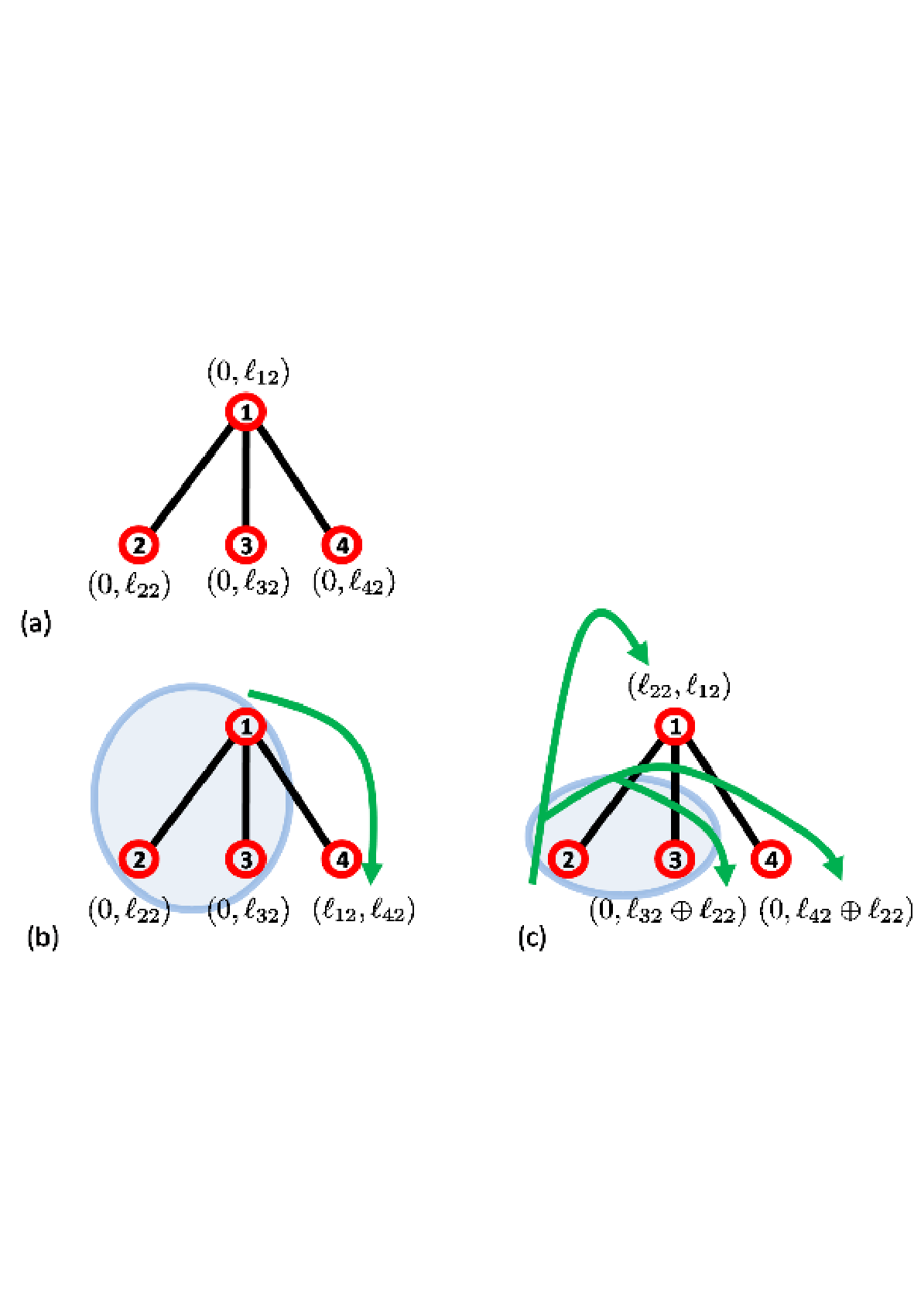}}
  \caption{\label{fig:GHZMexample}
(Color
 online) Equivalence of labeled graph states depicting bit dependence for four players.
(a)~The initial four party encoded graph state corresponds to $\vec{\ell}_{i\star}=(0,l_{i2})$.
(b)~Equivalent state with~$\vec{\ell}_{1\star}=(0,0)$, $\vec{\ell}_{2\star}=(0,l_{22})$,
	$\vec{\ell}_{3\star}=(0,l_{32})$, and $\vec{\ell}_{4\star}=(l_{12},l_{42})$ (see Eqn.~(\ref{eq:4GHZMdepEGb})):
	bit $\ell_{12}$ has been shuffled from vertex~1 to vertex~4, which is indicated by the arrow in the figure.
(c)~Equivalent state with bit $\ell_{22}$ shuffled to the other parties:
	$\vec{\ell}_{1\star}=(\ell_{22},l_{12})$, $\vec{\ell}_{2\star}=(0,0)$,
	$\vec{\ell}_{3\star}=(0,\ell_{22}\oplus l_{32})$, and $\vec{\ell}_{4\star}=(0,l_{22}\oplus l_{42})$
	as indicated by arrows and labels (see Eqn.~(\ref{eq:4GHZMdepEGc)})).
	NOTE: vertex $\mathsf{v}_i$ without label is equivalent to setting $\vec{\ell}_{i\star}=(0,0)$. }
\end{figure}

Graphically we ascertain the dependancy of a set of players $\mathsf{V}'$ on the information by shuffling bits between neighbours: that is, exploiting
equivalences  between labeled graph states with different labels~$\vec{\ell}$.
If a label does not occur on one of the vertices in the set $\mathsf{V}'$, that set has no dependency on it. This is obvious because the label would then represent a local unitary operation outside the set $\mathsf{V}'$, which cannot affect the reduced density matrix $\rho_{\mathsf{V}'\vec{\ell}}$. We thus look for equivalences between labeled graph states where labels can be shuffled outside $\mathsf{V}'$ as much as possible.
Of course different $\vec{\ell}_{\star 2}$-labeled graph states,
equivalently encoded graph states,
must be inequivalent because of the orthonormality relation~(\ref{eq:Gorthonormality});
indeed this is why we need the~$\vec{\ell}_{\star 1}$ labels.

Then the next step is to ascertain if the desired information can be accessed by~$\mathsf{V}'$, which
is effected by a measurement protocol corresponding to
appropriate combinations of $K_i$~(\ref{eq:eigenoperators}),
then exploiting the eigenequation relations~(\ref{eq:eigenequations}).

Fortunately the graph readily reveals the appropriate way to do this.
Before giving the general rules, we begin with a four-qubit example.

\begin{example}
\label{ex:4rules}
	A dealer distributes the encoded graph state~(\ref{eq:4GHZMdependencyEG})
	to all four players
		$\mathsf{V}=\{\mathsf{v}_1,\mathsf{v}_2,\mathsf{v}_3,\mathsf{v}_4\}$
	depicted in Fig.~\ref{fig:GHZMexample}(a).
	In this example we show that, if the three players in~$\mathsf{V}'=\{\mathsf{v}_1,\mathsf{v}_2,\mathsf{v}_3\}$ collaborate,
	they learn two bits $(\ell_{22},\ell_{32})$ of the four-bit encoding and are entirely denied knowledge of
	two other bits $(\ell_{12},\ell_{42})$, whether they use shared quantum channels or only LOCC.
	We also see that just two players  $\mathsf{V}''=\{\mathsf{v}_2,\mathsf{v}_3\}$ can (and can only) acquire $\ell_{22} \oplus \ell_{32}$ via LOCC,
	where~$\oplus$ is summation $\mod 2$.
\end{example}

Acting upon the encoded graph state~(\ref{eq:4GHZMdependencyEG})
  by the fourth stabilizer of
Eq.~(\ref{eq:4GHZMdependencyEG eigen-ops}) yields
\begin{align}
\label{eq:4GHZMdepEGb}
	K_4^{\ocircle \ell_{12}}|\mathcal{G}_{\vec{\ell_{\star 2}}}\rangle
		= & \openone_1 \otimes Z_2^{\ell_{22}} \otimes Z_3^{\ell_{32}}
			\otimes X_4^{\ell_{12}}Z_4^{\ell_{42}}\left|\mathcal{G}\right\rangle  \nonumber \\
		= &  \left|\mathcal{G}_{\vec{\ell}=(0,0,0,
			\ell_{22},0,\ell_{32},\ell_{12},\ell_{42})}\right\rangle \nonumber \\
        =& (-1)^{\ell_{12}\ell_{42}} |\mathcal{G}_{\vec{\ell_{\star 2}}}\rangle.
\end{align}
The player subset $\mathsf{V}'=\{\mathsf{v}_1,\mathsf{v}_2,\mathsf{v}_3\}$
is thus independent of $\ell_{12}$ because local unitary dependence on $\ell_{12}$
has been shuffled to player~$\mathsf{v}_4$'s two-bit label~$\vec{\ell}_{4\star}$.
In Fig.~\ref{fig:GHZMexample}(b), this $\ell_{12}$ bit shuffle is depicted
as a green arrow from vertex~$\mathsf{v}_1$ to vertex~$\mathsf{v}_4$.
Vertex~$\mathsf{v}_1$ is unlabeled because its bit value is now zero,
and vertex~$\mathsf{v}_4$ now has a two-bit label $\vec{\ell}_{4\star}=(\ell_{12},\ell_{42})$.
The graph states in Figs.~\ref{fig:GHZMexample}(a) and~\ref{fig:GHZMexample}(b)
are thus equivalent up to (an irrelevant) global phase.

Players in~$\mathsf{V}'$ share a reduced state that is dependent on
bits~$\ell_{22},\ell_{32}$.
In fact they can perform measurements
corresponding to $K_2^\ocircle$ and~$K_3^\ocircle$ to learn
these two bits according to~(\ref{eq:eigenequations}).
With shared quantum channels, there is no problem in performing such a
joint measurement as $K_2^\ocircle$ and~$K_3^\ocircle$ commute and have
support only on operations by players in $\mathsf{V}'$.
However quantum channels are not required: the players can learn
$(\ell_{22},\ell_{32})$ by LOCC because  the stabilizers commute locally.
In contradistinction a failure of local commutation would imply that
these bits could not be learned by LOCC.

In fact a smaller set of players can access information.
Consider the set of players $\mathsf{V}''=\{\mathsf{v}_2,\mathsf{v}_3\}$.
Applying eigenoperator~$K_1^{\ocircle {\ell_{22}}}$ yields
\begin{align}
\label{eq:4GHZMdepEGc)}
	K_1^{\ocircle {\ell_{22}}} \left|\mathcal{G}_{\vec{\ell}_{\star 2}}\right\rangle
		=&  X^{\ell_{22}}Z^{\ell_{12}} \otimes \openone \nonumber \\ &
			\otimes Z^{\ell_{22}+\ell_{32}} \otimes Z^{\ell_{22} + \ell_{42}}|\mathcal{G}\rangle
				\nonumber	\\
		= & \left|\mathcal{G}_{\vec{\ell}
			=(\ell_{22},\ell_{12},0,0,0,\ell_{22}\oplus \ell_{32},0,\ell_{22}\oplus \ell_{42})}\right\rangle \nonumber \\
=& (-1)^{\ell_{12}\ell_{22}} \left|\mathcal{G}_{\vec{\ell}_{\star 2}}\right\rangle
\end{align}
This implies that the reduced density matrix only depends on
$\ell_{22} \oplus \ell_{32}$. This information can be accessed by measuring
\begin{equation}
\label{eq:K2K3}
	K_2^\ocircle K_3^\ocircle= \openone \otimes X \otimes X \otimes \openone.
\end{equation}
This can be done by LOCC. This shuffling is illustrated in Fig.~\ref{fig:GHZMexample}(c);
hence all graph states in Figs.~\ref{fig:GHZMexample}(a) to~\ref{fig:GHZMexample}(c) are equivalent.

\begin{remark}
Stabilizer operators $K_i^\ocircle$ (and products thereof) are not directly measured by the players under LOCC;
rather players locally measure in the~$X_i$, $Y_i=\text{i}X_iZ_i$, or $Z_i$ bases
to obtain one-bit  outcomes $s_i^X$, $s_i^Y$, and~$s_i^Z$, respectively,
where $s_i^\alpha$ is assigned the value~$0$ if the
measurement outcome is~$1$, and~$1$ if the measurement outcome is~$-1$.

Then the bit value outcome for the measurement of $K_2^\ocircle$ and~$K_3^\ocircle$
in Example~\ref{ex:4rules} is
\begin{equation}
	\ell_{22}=s_1^Z \oplus s_2^X, \,
	\ell_{32}=s_1^Z \oplus s_3^X,
\end{equation}
respectively.
Henceforth when we say players can locally measure an operator,
we mean measurement in the sense that local Pauli measurements can be performed
and resultant bit outcomes combined to learn the same as if nonlocal
stabilizer measurements were performed.
\end{remark}

For Example~\ref{ex:4rules}, each player's qubit corresponded to a $\ocircle$ vertex.
Extending to $\square$ vertices is straightforward
by using the $K_i^{\square}$ stabilizers~(\ref{eq:eigenoperators}).
Specifically if a vertex is modified by changing its graph representation
from a~$\ocircle$ to a~$\square$, in the stabilizers an~$X$ is changed to a~$Y$
and we have a corresponding change to the bit transfer.

\begin{example}
\label{ex:4GHZMsquare}
	For Examples~\ref{ex:4GHZM} and~\ref{ex:4rules}
	let the~$\ocircle$ representation of vertex~$\mathsf{v}_1$
	be changed to a~$\square$ in Fig.~\ref{fig:GHZMexample}(a).
	Thus, in Fig.~\ref{fig:GHZMexample}(c),
	the resultant bit pair on vertex~$\mathsf{v}_1$ is $\vec{\ell}_{1\star}=(\ell_{22},\ell_{12}\oplus \ell_{22})$
	instead of~$\vec{\ell}_{1\star}=(\ell_{22},\ell_{12})$ because modifying $X\mapsto Y=\text{i}XZ$ in the revised
	stabilizer $K_1^\ocircle \mapsto K_1^\square$ yields an extra $Z$
	operation leading to~$\oplus 1$ on the second bit (via the appropriate change to Eq,~(\ref{eq:4GHZMdepEGc)})). The others remain the same $\vec{\ell}_{3 \star}=(0,\ell_2 \oplus \ell_3)$, $\vec{\ell}_{4\star}=(0, \ell_2 \oplus \ell_4)$.
\end{example}

These examples for four qubits are generalizable to $n$ qubits with
$\ocircle$ or $\square$ vertices. Here are the general principles for accessibility and
dependence of information in an encoded graph state.

\begin{proposition}
\label{prop:DepAcc}
The following principles hold for encoded graphs states $|\mathcal{G}_{\vec{\ell}_{\star 2}}\rangle$.
\begin{itemize}
\item [P1]
	The encoded graph state $|\mathcal{G}_{\vec{\ell}_{\star 2}}\rangle$
	with one-bit label~$l_{i2}$ on vertex~$\mathsf{v}_i$
	is equivalent to the labeled graph state $|\mathcal{G}_{\vec{\ell}}\rangle$
	where vertex~$\mathsf{v}_i$ is relabeled~$0$ and
	a neighbouring vertex~$\mathsf{v}_j$ is relabeled either
	$\vec{\ell}_{j\star}=(\ell_{i2},\ell_{j2})$ or $\vec{\ell}_{j \star}=(\ell_{i2},\ell_{j2}\oplus \ell_{i2})$, depending on whether
	vertex~$\mathsf{v}_i$ is of the type~$\ocircle$ or~$\square$, respectively.
	The remaining neighbours of $\mathsf{v}_j$, $\mathsf{v}_k \in N_i$
	are relabeled $(0,\ell_{k2} \oplus \ell_{i2})$.
	In this way any neighbour of $\mathsf{v}_i$ can shuffle the dependency of $\ell_{i2}$
	to itself and to its remaining neighbours.
	Bit shuffling is depicted as green arrows
	and shows how dependence of the reduced state
	on bit labels can be transferred from one player to another.
\item [P2]
\label{Rule: Access i}
	The bit value $\ell_{i2}$ is accessible by players in set~$\mathsf{V}'$ if and only if
	$\mathsf{v}_i$ and all its neighbours~$N_i$
	are in the set~$\mathsf{V}'$.
\item [P3]
	If the set $\mathsf{V}'$ comprises an even number of players, and all their external neighbours
	(outside the set $\mathsf{V}'$) are neighbours of each and every player in $\mathsf{V}'$, then
	the sum (mod 2) of all their one-bit labels~$\{\ell_{i2}\}$ is accessible, i.e.\ they can collaborate to find
	$$\ell_{i2} \oplus \ell_{j2} ... \oplus \ell_{m2},\;\mathsf{v}_i, \mathsf{v}_j,...,\mathsf{v}_m \in \mathsf{V}' .$$
\item [P4]
	If~P2 holds, and players in the set share quantum channels, all information
	encoded in one-bit labels is accessible. If the set shares only classical channels,
	and P2 holds, then information accessibility is limited because
	neighbours $\mathsf{v}_i$ and~$\mathsf{v}_j$ in the set can only learn either $\ell_{i2}$ or $\ell_{j2}$ but not both,
	but $\mathsf{v}_i$ and~$\mathsf{v}_j$ in the set can learn both~$\ell_{i2}$
	and~$\ell_{j2}$ if they are not immediate neighbours.
\end{itemize}
NOTE: The bit shuffling in P1 is not active; rather it recognizes equivalence between labeled graph states.
\end{proposition}
\begin{proof}
All proofs are simply derived from eigenequations~(\ref{eq:eigenoperators}) and~(\ref{eq:eigenequations}), which shuffles bits between players, thereby revealing
actual dependence of reduced states on given bit labels~$\{\ell_{i2}\}$.

For~P1, suppose vertex~$\mathsf{v}_i$ has a neighbouring vertex~$\mathsf{v}_j$ which is either
a~$\ocircle$ or a~$\square$.
Then
\begin{widetext}
\begin{align}
|\mathcal{G}_{\vec{\ell}_{\star 2}}\rangle
	=& (-1)^{\ell_{i2}\ell_{j2}}K_j^{\ocircle \ell_{i2}}|\mathcal{G}_{\vec{\ell}_{\star 2}}\rangle
	= (-1)^{\ell_{i2}\ell_{j2}} \openone_i \otimes Z_j^{\ell_{j2}} X_j^{\ell_{i2}}
			\otimes_{(\mathsf{v}_k,\mathsf{v}_j)\in\mathsf{E}}Z_k^{(\ell_{k2} + \ell_{j2})}
			\otimes_{m\neq i,j ; (\mathsf{v}_m,\mathsf{v}_j)\notin\mathsf{E}} Z_m^{l_{m2}}|\mathcal{G}\rangle,
				\nonumber	\\
	|\mathcal{G}_{\vec{\ell}_{\star 2}}\rangle
		=& (-1)^{\ell_{i2}\ell_{j2}}K_j^{\square \ell_{i2}}|\mathcal{G}_{\vec{\ell}_{\star 2}}\rangle
		= (-1)^{\ell_{i2}\ell_{j2}} i \openone_i \otimes Z_j^{l_{i2} \oplus \ell_{j2}} X_j^{\ell_{i2}} \otimes_{(\mathsf{v}_k,\mathsf{v}_j)\in\mathsf{E}}Z_k^{(\ell_{k2} + \ell_{j2})}
			\otimes_{m\neq i,j ; (\mathsf{v}_m,\mathsf{v}_j)\notin\mathsf{E}} Z_m^{\ell_{m2}}|\mathcal{G}\rangle,
\end{align}
\end{widetext}
respectively.
Evidently dependency on $\ell_{i2}$ is shuffled from neighbour~$\mathsf{v}_i$ to~$\mathsf{v}_j$,
and $\mathsf{v}_j$'s other neighbours are as described in P1.

For~P2, players obtain~$\ell_{i2}$ by measuring $K_i^{\ocircle,\square}$ of~(\ref{eq:eigenoperators})
and~(\ref{eq:eigenequations}). Because
$K_i^{\ocircle, \square}$ is nontrivial on~$i$ and also on~$N_i$,
by~(\ref{eq:eigenoperators}), player~$i$
and all of~$N_i$ must collaborate to measure~$\ell_{i2}$, and this collaboration is sufficient.

For~P3 we start by looking at the case $\mathsf{V}'$ contains only two players. By~(\ref{eq:eigenoperators}) if all neighbours
of pair $\mathsf{v}_i,\mathsf{v}_j\in \mathsf{V}'$ are either in the set, or shared outside the set,
$K_i^{\ocircle, \square} \cdot K_j^{\ocircle, \square}$ is nontrivial only inside the set, and thus
can be measured by the members of the set. This is true since all the~$Z$ from the~$K_i^{\ocircle, \square}$ cancel with the~$Z$ from  $K_j^{\ocircle, \square}$ on their common neighbours.
Applying $K_i^{\ocircle, \square} \cdot K_j^{\ocircle, \square}$ to~(\ref{eq:eigenequations}), we see that this yields $\ell_{i2}
\oplus \ell_{j2}$. Similarly this holds for any even number of players.

For the comparison of LOCC to the fully quantum case of~P4, by the above logic if properties
P2 and~P3 hold, then the dependency on the associated observables are
nontrivial only within the set. As all operators $K_i^{\ocircle, \square}$
commute, all observables commute
in the nontrivial set, and hence can be measured by global operations.

For LOCC, it is clear from~(\ref{eq:eigenoperators})
that, if vertices~$\mathsf{v}_i$ and~$\mathsf{v}_j$ are neighbours, then~$K_i^{\ocircle, \square}$
and~$K_j^{\ocircle, \square}$ share different
local Pauli operators on $i$ and~$j$ hence cannot be simultaneously
measured locally;
therefore, even if P2 holds, one must choose one measurement or the other.
If vertices~$\mathsf{v}_i$ and~$\mathsf{v}_j$ are not neighbours,
they are indeed amenable to simultaneous local measurements. Similar statements can be made when P3 holds, but are more complicated graphically and must be taken case by case.
\end{proof}

\subsection{Local $Z$ and~$Y$ measurements on encoded graph states}
\label{subsec:measurements}

In our protocols encoded graph states with both $\square$ and $\ocircle$ vertices arise as
a result of performing either~$Y$ or~$Z$ measurements on qubits
identified by vertices of the labeled graph with only circular vertices.
Local Pauli measurements on graph states are described by a
simple set of rules, which readily yield the resultant states~\cite{HDERNB06}.
These rules were originally stated in terms of graph
states up to local unitary operations. For our purposes we would like to
incorporate also these local unitary operations into the labeled graph states. We now
represent these rules exactly graphically for the~$Z$ and~$Y$
measurements, using the labeled graphs above, with no local unitary
ambiguity.

If a~$Z$ measurement is made on vertex~$\mathsf{v}_i$ of an encoded graph state with circular vertices,
the resultant state corresponds to a labeled graph state of the
original graph but with vertex $\mathsf{v}_i$ and its edges deleted.
For outcome $s_i^Z$, the labels of all vertices in~$N_i$ change to
$\vec{\ell}_j\star \mapsto(0,\ell_{j2}\oplus s_i^Z)$.
We denote the resultant graph by $g$.

\begin{figure}
\scalebox{0.45}{\includegraphics*[0.5cm,7.5cm][21cm,22cm]{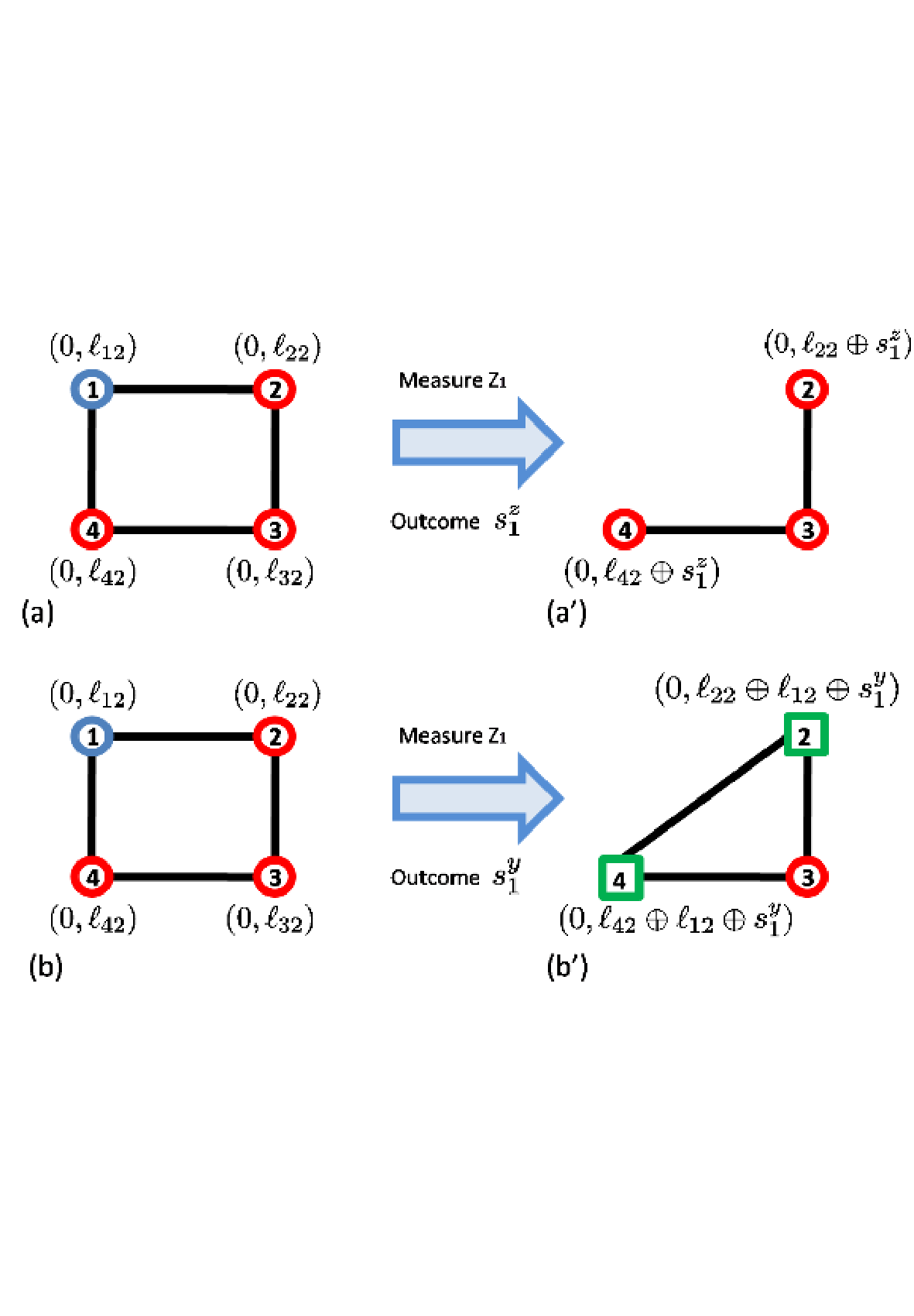}}
  \caption{\label{fig:measurements on a square}
	(Color
 online) When Pauli $Z$ and~$Y$ measurements are made on vertex $\mathsf{v}_1$, graphs~(a)
	and~(b) metamorphose into~(a$'$) and~(b$'$), respectively.
	Outcomes $S_i^{\alpha}=0,1$ correspond to measurement eigenvalues $-1,+1$ respectively, and 	are incorporated into the labels in~(a$'$) and~(b$'$).}
\end{figure}

If a~$Y$ measurement is made on vertex~$\mathsf{v}_i$ of an encoded graph state with circular vertices,
the resultant state is a labeled graph state
obtained by applying the following four steps to the original graph;
the new graph is denoted $g'$.
\begin{itemize}
\item [S1] Perform local complementation on~$\mathsf{v}_i$: all existing edges between elements in $N_i$ are removed, and where they did not exist, they are added.
\item [S2] Each $\ocircle$ vertex in $N_i$
	is changed to $\square$.
\item [S3] For outcome $s_i^Y$, the labels on the neighbours of $i$ change to $\ell_{j \star} =(0,\ell_{j2}) \rightarrow (0,\ell_{j2}\oplus \ell_{i2} \oplus s_i^Y)$
\item [S4] Remove vertex $i$ and all its edges.
\end{itemize}
\begin{example}
\label{ex:ZY}
	A~$Z$ and a~$Y$ measurement on vertex $\mathsf{v}_1$ of a square graph state with circular vertices
	changes the graphs as depicted in Fig.~\ref{fig:measurements on a square}.
\end{example}

These rules can easily be understood by the fact that any encoded graph state with labels $\ell_{i2}=0$ $\forall i$
can be written in the form
\begin{align}
\label{eq:conjugate meas}
    |\mathcal{G}_{\left(\vec{0}\right)}\rangle
    	=& \frac{1}{\sqrt2} \bigg( |0\rangle_1 |g_{(0\cdots 0)}\rangle_{2,\ldots,n} +
        |1\rangle_1 |g_{(\underbrace{1\cdots 1}_{\in N_1}0\cdots
            0)}\rangle_{2,\ldots,n} \bigg) \nonumber \\
        =& \frac{\text{e}^{i\pi/4}}{\sqrt2}
        		\bigg( |0'\rangle_1 |g'_{(0\cdots 0})\rangle_{2,\ldots,n}
			\nonumber	\\	&
		- i |1'\rangle_1 |g'_{(\underbrace{1\cdots 1}_{\in N_1}0
		\cdots 0)}\rangle_{2,\ldots,n} \bigg),
\end{align}
with $\{ |0\rangle , |1\rangle\}$ and
$$\{|0'\rangle=1/\sqrt{2}(|0\rangle-i|1\rangle), |1'\rangle =1/\sqrt{2} (|0\rangle+i|1\rangle)\}$$
the eigenstates of the~$Z$ and~$Y$ Pauli operators, and the corresponding
eigenvalues are $+1$ and~$-1$, respectively.
Adding labels $\vec{\ell}_{\star 2}$ indicates the application of local $Z$s. This commutes with measuring Pauli $Z$, so the labels carry forward only with the addition of the correction from the measurement itself, e.g. Fig.~\ref{fig:measurements on a square}(a),(a'). However, local $Z$ does not commute with a $Y$ measurement, hence the bit $\ell_{12}$ carries on to the measured graph (S3), e.g. Fig.~\ref{fig:measurements on a square}(b),(b').


The following definition follows naturally.
\begin{definition}
	Given a graph~$\mathcal{G}$ with only circular vertices, and associated encoded graph state
	$|\mathcal{G}_{\vec{\ell}_{\star 2}}\rangle$, the graph of the state given by measuring Pauli $Y$ on vertex
	$i\in \{1,\ldots,n\}$ is called the \emph{conjugate~graph} denoted $g'$, with \emph{conjugate~graph~state} denoted
	$|g'\rangle_{1,\ldots,n/i}$ and can be found explicitly by following the rules
\begin{itemize}
\item [R1] Perform local complementation on~$\mathsf{v}_i$.
\item [R2] Replace all neighbours of~$i$ with $\square$ vertices.
\item [R3] Remove~$\mathsf{v}_i$ and all its edges.
\end{itemize}
\end{definition}

It can easily be shown that encoded conjugate graph states can be written as
\begin{align}
\label{eq:conjugate graph states}
\left|g'_{(0\cdots 0)}\right\rangle =& \frac{e^{-i\pi/4}}{\sqrt2} \bigg( |g_{(0\cdots 0)}\rangle +
        |g_{(\underbrace{1\cdots 1}_{\in N_1}0\cdots
            0)}\rangle \bigg) \nonumber \\
\bigg|g'_{(\underbrace{1\cdots 1}_{\in N_1}0
            \cdots 0)}\bigg\rangle =& \frac{e^{-i\pi/4}}{\sqrt2} \bigg( |g_{(0\cdots 0)}\rangle -i
        |g_{(\underbrace{1\cdots 1}_{\in N_1}0\cdots
            0)}\rangle \bigg)
\end{align}
so are effectively conjugate to the encoded graph states
$$\{\left|g_{(0\cdots 0)}\right\rangle, |g_{(\underbrace{1\cdots 1}_{\in N_1}0\cdots 0)}\rangle \}.$$

The following definitions are useful for Sec. \ref{sec: embedded}.


\begin{definition}
	The $n$GHZM graph $\mathcal{G}_{n\text{GHZM}_i}$
	is \emph{embedded} in~$\mathcal{G}$ if~$\mathcal{G}_{n\text{GHZM}_i}\subset\mathcal{G}$ and
	$N_i\cap \mathcal{G}\backslash\mathcal{G}_{n\text{GHZM}_i}=\emptyset$.
\end{definition}
\begin{definition}
	The \emph{$n$GHZM graph state} is \emph{embedded}
	into state $|\mathcal{G}\rangle$ if~$\mathcal{G}_{n\text{GHZM}_i}$
	is embedded in $\mathcal{G}$.
\end{definition}

We now have the necessary formalism to develop the various secret sharing
protocols.

\section{Secret Sharing Protocols}
\label{sec:protocols}

In this section we introduce our three secret sharing protocols
beginning with {\sf CC} for sharing classical secrets in Subsec.~\ref{subsec:direct}.
Using our graph state approach, we construct solutions for
each of $(n,n)$, $(3,4)$ and~$(3,5)$ threshold secret sharing.
These results provide a foundation for our subsequent protocols:
{\sf CQ} in Subsec.~\ref{subsec:CQ} and {\sf QQ} in Subsec.~\ref{subsec:QQ}.

\subsection{{\sf CC}}
\label{subsec:direct}

In the {\sf CC} protocol,
classical information is directly encoded into the bit string~$\vec{\ell}_{\star 2}$,
i.e.~onto the encoded graph state~$\{|\mathcal{G}_{\vec{\ell}_{\star 2}}\rangle \}$ in a manner
that enables certain sets of collaborating players to access concealed information
according to the principles of the
secret sharing protocol.

Mathematically our protocol is similar to Gottesman's~\cite{Got00} wherein
stabilizers are used to share classical secrets in quantum states. A difference between
our scheme and Gottesman's is that he uses mixed high-dimensional states
whereas our graph state approach enables use of pure states on qubits.
Another difference with Gottesman's approach is that our method is not as general:
our scheme works for a limited number of $(k,n)$ scenarios whereas
Gottesman's method works for any $(k,n)$.
Although our scheme is more restrictive, it is applicable to a restricted set of
cases and importantly provides a foundation for the protocols discussed in subsequent subsections.

The {\sf CC} protocol steps are as follows:
\begin{itemize}
\item [{\sf CC}1]The dealer encodes the
secret bit string~$S$ into the classical information $\vec{\ell}_{\star 2}$ and encodes this onto the
$n$-qubit state $\{|\mathcal{G}_{\vec{\ell}_{\star 2}}\rangle \}$.
\item [{\sf CC}2] The dealer sends each player one qubit.
\item [{\sf CC}3] An authorized set of players
	accesses the secret bit by measuring appropriate stabilizer operators
	(prescribed explicitly in each case).
\end{itemize}
Here we analyze the ability of sets of players to access this
information for two cases: using LOCC and using full quantum communication (FQC).
We will see that each graph represents a different sharing structure, and that
the rules of access are simple in graph terms, and show that some
graphs can be used as secret sharing schemes as set out in the introduction.
In particular we present schemes for $(n,n)$, $(3,4)$ and~$(3,5)$.

\begin{figure}
   \scalebox{0.46}{\includegraphics*[0.5cm,11cm][21.6cm,19cm]{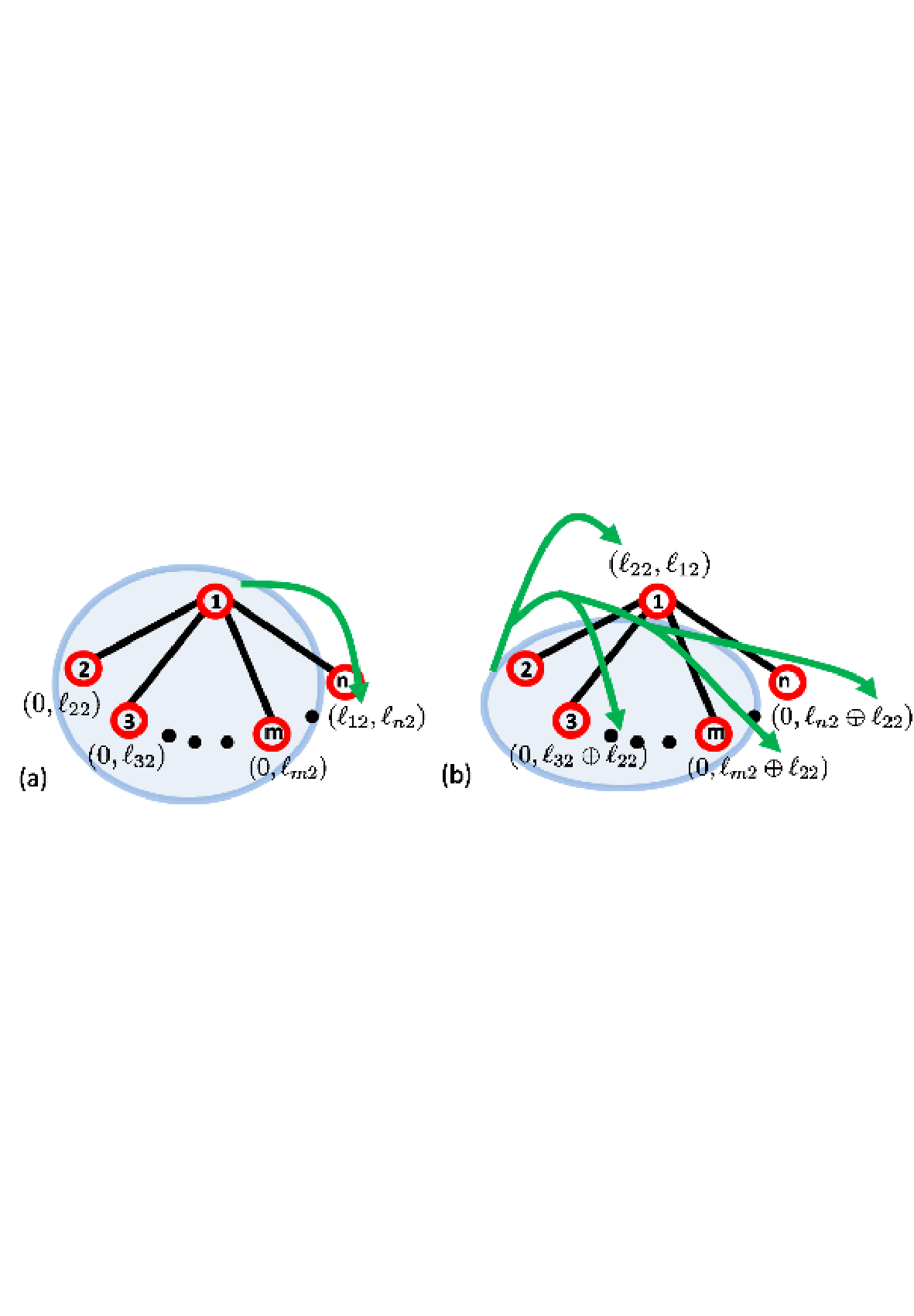}}
  \caption{\label{fig:GHZ Dependency}
    (Color
 online) Distribution and access of information
    $\{l_{12},\ldots,l_{n2}\}$ for the n-GHZM state. The green arrows represent uses of dependency Property P1.
    In~(a) the dependency on $l_{12}$ is taken out of the set, and in~(b) the
    dependency on $l_{22}$ is shared across the whole set.
    This encoded graph state gives an $(n,n)$ secret sharing scheme by setting $l_{12}=S$ and
    the other bits can be set arbitrarily.}
\end{figure}

We now proceed to give explicit examples of how the dependence and access properties in
Sec.~\ref{subsec:properties} can be applied to give secret
sharing via the above protocol.
We begin by looking at the case of $n$GHZM$_i$ states.
Using Property P1, we can see in Fig.~\ref{fig:GHZ Dependency}(a) that
the player set $\{\mathsf{v}_1,\ldots,\mathsf{v}_m\}$ depends only on bits $\ell_{22},\ldots,\ell_{m2}$;
i.e., the dependency on $\ell_{12}$ has been removed.
From P2 it is possible for the same player set to recover all $\ell_{22}\ldots ,\ell_{m2}$, and,
by P4, this can be done by LOCC.

Explicitly this would be done by
player~$1$ measuring $Z_1$ and the remaining players $\{\mathsf{v}_2,\ldots,\mathsf{v}_m\}$ measuring $X_i$ and comparing results to measure $K_2^\ocircle,\ldots,K^\ocircle_m$, yielding $\ell_{22},\ldots,\ell_{m2}$ via
eigenequations~(\ref{eq:eigenequations}).
Hence, in this case, LOCC is sufficient for all possible information access.

Similarly Fig.~\ref{fig:GHZ Dependency}(b) shows how P1 implies that
the player subset $\{\mathsf{v}_2,\ldots,\mathsf{v}_m\}$ depends on bits
$$l_{22} \oplus \ell_{32}, \ell_{22} \oplus \ell_{42}, \ldots , \ell_{22} \oplus \ell_{m2}$$
only and, by~P3 and~P4, they can be accessed simultaneously by LOCC alone.
Explicitly this
would be done by all players measuring $X_i$.
The players in this set then compare results to measure
\begin{equation}
	K_2^\ocircle \cdot K_3^\ocircle, K_2^\ocircle \cdot K_4^\ocircle, \ldots K_2^\ocircle \cdot K_m^\ocircle.
\end{equation}
Only in the case that all players collaborate would there be a difference between the LOCC and FQ cases;
in this case P4 implies that,
by LOCC, the players can either access $\ell_{22},\ldots ,\ell_{n2}$ \emph{or} $\ell_{12}$.
Obviously, for FQC, all information is simultaneously accessible.

We see above that bit $\ell_{12}$ is only accessible when all players act together.
By choosing the secret bit
$S$ as $S=\ell_{12}$ and fixing all other $\ell_{i2}$ arbitrarily, we obtain the following
proposition.
\begin{proposition}
\label{prop:(n,n)}
An encoded $n$GHZM state with $\ell_{12}=S$ and all other $\ell_{i2}$ set to zero
enables a~$(n,n)$ direct secret sharing scheme that can be accessed via LOCC.
\end{proposition}
\noindent
This case is depicted in Fig.~\ref{fig:GHZ Dependency}.
\begin{figure*}
\scalebox{0.7}{\includegraphics*[0.5cm,12cm][21cm,18cm]{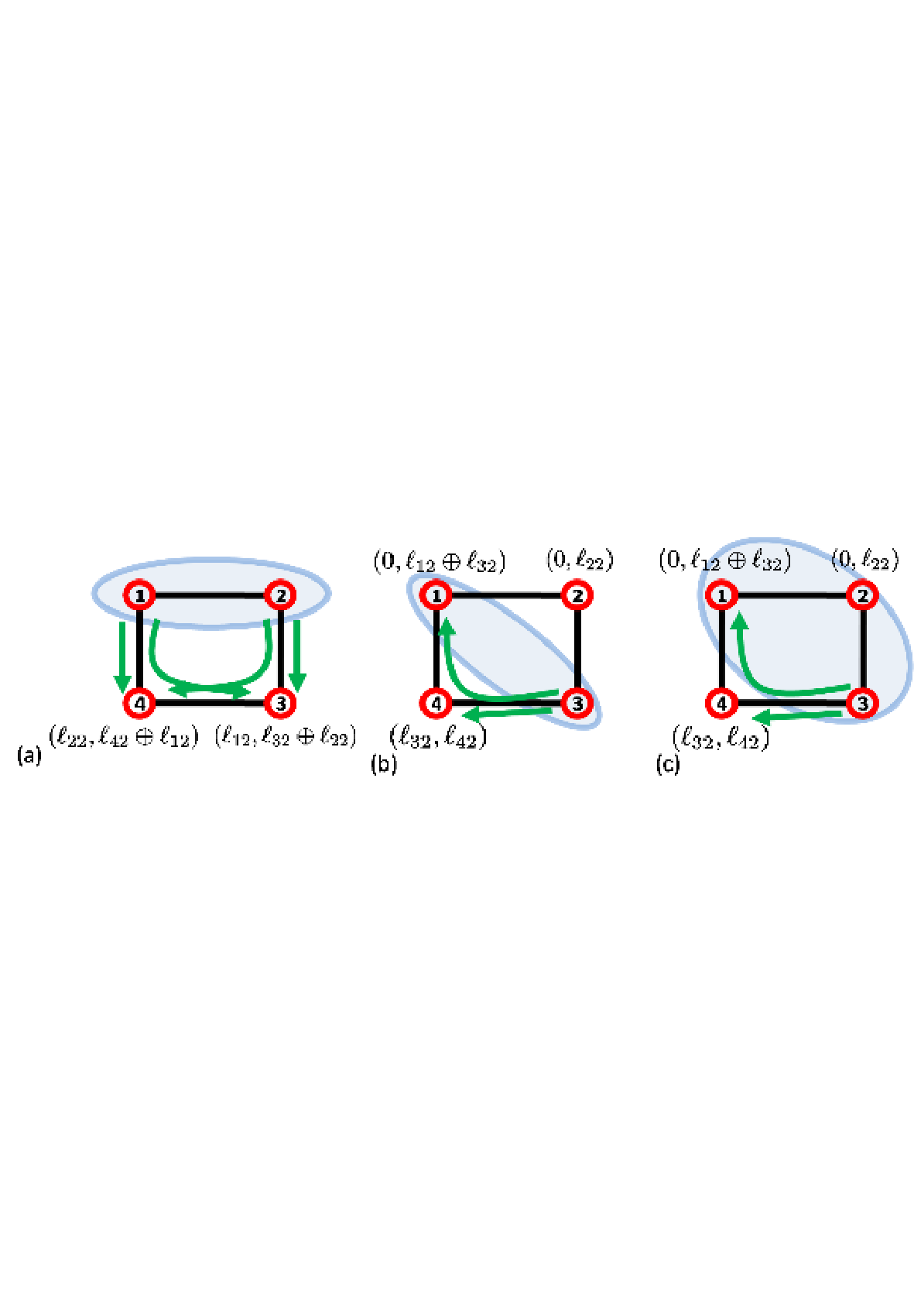}}
  \caption{\label{fig:square}
    (Color
 online) Application of P1 to the square encoded graph state. This
    graph gives a~$(3,4)$ secret sharing scheme setting $\ell_1=~\ell_2=~\ell_3=~\ell_4=~S$ (see proposition \ref{prop:(3,4)}).}
\end{figure*}

Now we see that $(3,4)$ threshold secret sharing is attainable with encoded graph states.
\begin{proposition}
\label{prop:(3,4)}
The $4$-qubit ring encoded graph state with $\ell_{12}=\ell_{22}=\ell_{32}=\ell_{42}=S$ enables a~$(3,4)$ direct secret sharing
scheme, and LOCC between players suffices for decoding.
\end{proposition}
\begin{proof}
The $4$-qubit ring state is depicted in~Fig.~\ref{fig:square}.
We see in Fig.~\ref{fig:square}(a) that P1 shows that pairs
$\{\mathsf{v}_1,\mathsf{v}_2\}$, $\{\mathsf{v}_2,\mathsf{v}_3\}$, $\{\mathsf{v}_3,\mathsf{v}_4\}$ and~$\{\mathsf{v}_1,\mathsf{v}_4\}$ are denied any information whatsoever.

Figure~\ref{fig:square}(b) shows that the player pair $\{\mathsf{v}_1,\mathsf{v}_3\}$
can obtain, and moreover can only obtain, $\ell_{12} \oplus \ell_{32}$.
Similarly $\{\mathsf{v}_2,\mathsf{v}_4\}$ obtain,
and only obtain, $\ell_{22} \oplus \ell_{42}$.
Fig.~\ref{fig:square}(c) reveals that players~$\{\mathsf{v}_1,\mathsf{v}_2,\mathsf{v}_3\}$ obtain, and only obtain, $\ell_{22}$ and/or
(by using FQC/LOCC) $\ell_{12} \oplus \ell_{32}$.

Similarly, players $\{\mathsf{v}_2,\mathsf{v}_3,\mathsf{v}_4\}$ obtain, and only obtain, $\ell_{32}$ and/or (via FQC/LOCC) $\ell_{22} \oplus \ell_{42}$.
The remaining triplets obtain, and only obtain,
$\ell_{12}$ and/or (via FQC/LOCC) $\ell_{22} \oplus \ell_{42}$ in the case of $\{\mathsf{v}_1,\mathsf{v}_2,\mathsf{v}_4\}$,
and~$\ell_{42}$ and/or (FQC/LOCC) $\ell_{12} \oplus \ell_{32}$ in the case of $\{\mathsf{v}_1,\mathsf{v}_3,\mathsf{v}_4\}$.

We see that this graph gives a (3,4) threshold scheme by setting
\begin{equation}
	\ell_{12}=\ell_{22}=\ell_{32}=\ell_{42}=S.
\end{equation}
\end{proof}

\begin{figure*}
\scalebox{0.8}{\includegraphics*[0.5cm,12cm][21cm,17.5cm]{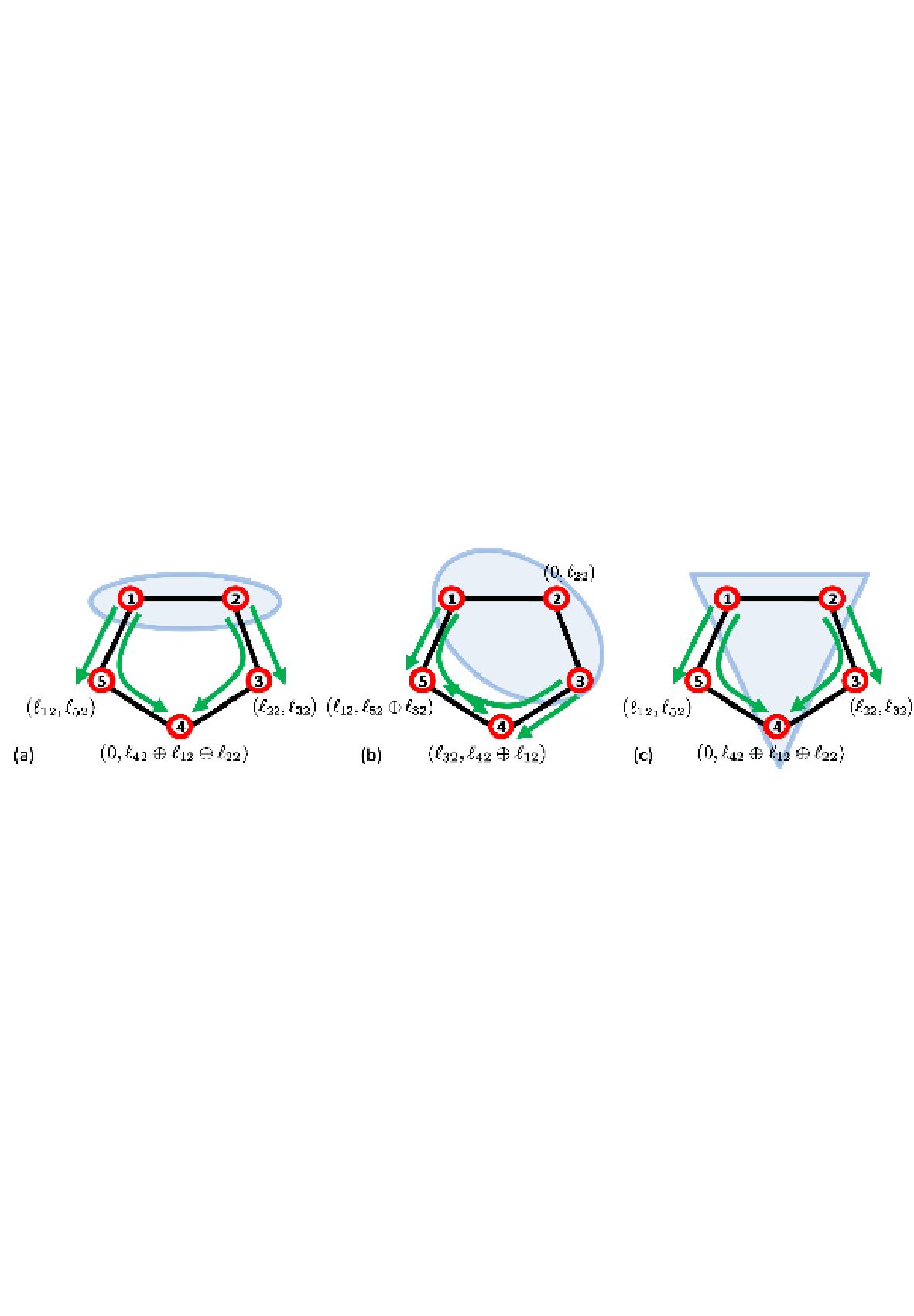}}
  \caption{\label{fig:pentagon}
    (Color
 online) Application of~P1 to the~$5$-qubit ring encoded graph state to obtain
    a~$(3,5)$ direct secret sharing scheme by setting $\ell_{12}=~\ell_{22}=~\ell_{32}=~\ell_{42}=~\ell_{52}=~S$.}
\end{figure*}

Our final result in this subsection concerns the~$(3,5)$ threshold scheme, given by a
$5$-qubit ring depicted in Fig.~\ref{fig:pentagon}.
\begin{proposition}
\label{prop:(3,5)}
	The five-qubit encoded RING graph state enables $(3,5)$ threshold secret sharing
	for the {\sf CC} protocol provided that $\ell_{12}=\ell_{22} = \ell_{32} = \ell_{42} = \ell_{52} = S$,
	with LOCC between players being sufficient for decoding.
\end{proposition}

\begin{proof}
Fig.~\ref{fig:pentagon}(a)
shows how P1 implies that players~$1$ and~$2$ cannot access any
information because the dependency can be taken away from them.
Similarly it is easy to see that no pair of players can access any information by application of~P1.
On the other hand if three cyclic neighbours $\{\mathsf{v}_i,\mathsf{v}_j,\mathsf{v}_k\}$ form a set,
P1 implies the set only depends on $\ell_{j2}$; this is clear
from the case for the player set $\{\mathsf{v}_1,\mathsf{v}_2,\mathsf{v}_3\}$ in Fig.~\ref{fig:pentagon}(b).
P2 implies the bit can be accessed, and P4 says this access can be achieved via LOCC.
Accessing would be accomplished by measuring $K_j^\ocircle$,
which is done by locally measuring $Z_i$, $X_j$, $Z_k$.

The remaining possible sets of three players are all similar to the~$T$
shape in Fig.~\ref{fig:pentagon}(c);
in this case the set $\{\mathsf{v}_i,\mathsf{v}_j,\mathsf{v}_k\}$
depends only on $\ell_{i2} \oplus \ell_{j2} \oplus \ell_{k2}$ by property P1.
This is seen for the example $\{\mathsf{v}_1,\mathsf{v}_2,\mathsf{v}_4\}$ in Fig.~\ref{fig:pentagon}(c). By properties P2 and~P4 the players can access the secret by LOCC.
Their procedure is effected by measuring $K_i^\ocircle \cdot K_j^\ocircle \cdot K_k^\ocircle$,
which is accomplished locally by measuring $Y_i$, $Y_j$, $X_k$.

Any set of more than three players contains at least one of these possibilities,
and the players can therefore access the same information.
By choosing to encode the bit secret~$S$ as
\begin{equation}
	\ell_{12}=\ell_{22}=\ell_{32}=\ell_{24}=\ell_{52}=S,
\end{equation}
the proposition is proved.
\end{proof}

Here we established three examples that use graph states for secret sharing.
These three examples demonstrate the simple properties of our approach,
and more examples of graph states for secret sharing could be found.
In the next two subsections, we extend these results and techniques to secret key distillation
and to sharing a quantum secret.

\subsection{\sf CQ}
\label{subsec:CQ}

In this subsection we address case~{\sf CQ}, which corresponds to the protocol expounded
by Hillery, Bu\v{z}ek, and Berthiaume (HBB) for sending a classical secret with insecure quantum channels between the dealer and players~\cite{HBB99}.
Actually this problem can instead be solved by combining the classical secret sharing
protocol of Shamir~\cite{Sha79} via standard quantum key distribution (QKD)~\cite{BB84}
to assure secure classical channels between the dealer and each of the players.
In the QKD approach, the dealer would first establish a random key with each of the players individually by using QKD. This would enable perfectly secure classical communication between the dealer and each player (as a one time pad), which can then be used to perform the Shamir protocol safely.
However, QKD may not afford the most efficient approach hence the interest in alternative schemes.

HBB~\cite{HBB99} combine secret sharing protocols with QKD:
the dealer distributes multipartite states to share a secret key,
which can only be distilled by or shared with authorized sets of players.
This key can then be used to send any classical message the dealer would wish to send to the authorized players, without any unauthorized parties having access to it.
The HBB approach can be more efficient in the number of uses of quantum channels between dealer and players,
which is comparable to the greater efficiency achieved by direct distillation of multipartite entanglement
rather than distillation of pairs of maximally entangled states
and complicated sequences of teleportations~\cite{MJPV99}).

Another benefit is that the HBB approach can be extended to sharing quantum secrets ({\sf QQ}) as we see
in the next subsection.
HBB~\cite{HBB99} created a classical threshold scheme for $(2,2)$ and~$(3,3)$ cases, and this
has been extended to the~$(n,n)$ case in~\cite{Xiao04}.

We note that, as an extension of HBB, our protocol shares the same drawbacks.
The HBB protocol, hence ours, protects against intercept-resend attacks
but is not guaranteed to be secure against other attacks~\cite{KKI99}.
Second, it only works for a noiseless channel:
In the case of a noisy channel, the HBB protocol can not establish a key
(as the noise will be mistaken for an eavesdropper).
This problem is likely not fatal, as approaches involving entanglement distillation or alternatives
could help to overcome the effects of noise in the channel~\cite{Chen04}.

Our protocol employs graph states to generalize the HBB scheme~\cite{HBB99}.
By using graph states our scheme could build on
current experimental research for generating graph states for
measurement-based quantum computing~\cite{WRRSWVAZ05}.
Further, encoding onto graph states allows us to use the properties developed Sec.~\ref{sec: graphstates} to generalize to other access structures.

As an example we show how this approach generalizes HBB's protocol to the~$(3,5)$ case, together with the~$(n,n)$ case.
Although other examples are not provided here,
the properties given in Sec.~\ref{subsec:properties} allow further exploration.
In addition, our graph state shows how the HBB protocol can be thought of as an extension to Gottesman's results~\cite{Got00},
in the sense that HBB's protocol can be regarded as a purification of Gottesman's.

\bigskip

The underlying concept for the protocol is simple. We use an entangled state as the channel between the dealer~$D$ and~$n$ players. This will be a graph state; hence we can always write
in the following way (c.f.\ Sec.~\ref{subsec:measurements})
\begin{align}
\label{eq:conjugate meas SS}
    |\mathcal{G}_{(\vec{0})}\rangle =& \frac{1}{\sqrt2} \Big( |0\rangle_D |
    		g_{(0\cdots 0)}\rangle_{1,\ldots,n}
			\nonumber	\\	&
	+|1\rangle_D |g_{(\underbrace{1\cdots 1}_{\in N_D}0\cdots
         			0)}\rangle_{1,\ldots,n}\Big) \nonumber \\
        =& \frac{\text{e}^{i\pi/4}}{\sqrt2}\Big( |0'\rangle_D |g'_{(0\cdots 0)}\rangle_{1,\ldots,n}
			\nonumber	\\	&
            - i |1'\rangle_D |g'_{(\underbrace{1\cdots 1}_{\in N_D}0
            \ldots 0)}\rangle_{1,\ldots,n} \Big).
\end{align}
The dealer measures randomly in either the~$\{|0\rangle_D,|1\rangle_D\}$ or the~$\{|0'\rangle_D,|1'\rangle_D\}$ basis.
The dealer's measurement outcomes provide the random key.

Through the entanglement, the players can access the key by making the correct measurement. If the dealer measures in $\{|0\rangle_D,|1\rangle_D\}$, the results are then correlated to the states
$$\{|g_{(0\cdots 0)}\rangle_{1,\ldots,n}, |g_{(\underbrace{1\cdots 1}_{\in N_D}0\cdots 0)}\rangle_{1,\ldots,n}\}$$ held by the players. If any set of players can discriminate these states perfectly, they can find out what the measurement result of the dealer was, and thereby obtain the random key.

The dealer may instead choose to measure in the~$\{|0'\rangle_D,|1'\rangle_D\}$ basis,
and then the player's qubits will end up in either of the associated conjugate states
$$\{|g'_{(0\cdots 0)}\rangle_{1,\ldots,n},
 |g'_{(\underbrace{1\cdots 1}_{\in N_D}0\cdots 0)}\rangle_{1,\ldots,n}\}$$
correlated to the measurement result.
The ability to discriminate (or in other words, to access the information $\vec{\ell}_{\star 2}$ and hence get the key) is then determined by the conjugate graph $g'$, which must be checked to have the correct access structure also (if we want to do secret sharing). We see that this works for the GHZ and the 5-qubit ring encoded graph states, but it does not work for the 4-qubit ring encoded graph state. This is why we cannot extend the~$(3,4)$ {\sf CC}
protocol to key distribution in this way.

In the case of only one player, this corresponds to Ekert's protocol~\cite{Eke91}.
In our case there is the added structure given by the graph~$g$.
Discriminating these states is the same as accessing either the information
$\vec{\ell}_{\star 2} = \vec{0}$ or $\vec{\ell}_{\star 2} = (1,\ldots,1,0,\ldots,0)$.
By choosing the right graph, only the authorized players can access the information, or discriminate the states and access this key. In this first instance we use exactly the graphs of the last protocols to do this. In this way these entangled states~(\ref{eq:conjugate meas SS}) are a purification of the protocols of the last section - taking a random choice of the two graph states (which can be viewed as a mixed state) to an entangled pure state by adding an auxiliary system, which is the dealer in this case.

The players also measure randomly to either discriminate states

$$\{|g_{(0\cdots 0)}\rangle_{1,\ldots,n},
|g_{(\underbrace{1\cdots 1}_{\in N_D}0\cdots 0)}\rangle_{1,\ldots,n}\},$$
or the conjugate states
$$\{|g'_{(0\cdots 0)}\rangle_{1,\ldots,n}, |g'_{(\underbrace{1\cdots 1}_{\in N_D}0\cdots 0)}\rangle_{1,\ldots,n}\}.$$
As seen in Sec.~\ref{subsec:measurements} these states are conjugate in the sense that discriminating the states
$$\{|g_{(0\cdots 0)}\rangle_{1,\ldots,n}, |g_{(\underbrace{1\cdots 1}_{\in N_D}0\cdots 0)}\rangle_{1,\ldots,n}\}$$
provides no information about whether we have
$$|g'_{(0\cdots 0)}\rangle_{1,\ldots,n}$$
or
$$|g'_{(\underbrace{1\cdots 1}_{\in N_D}0\cdots 0)}\rangle_{1,\ldots,n}.$$
Thus, the players cannot discriminate both bases simultaneously, so even if a set of players is authorized and can discriminate perfectly,  50\% of the time they will measure in a basis in which they gain no information about the dealer's measurement result, hence will not access the key.
To accommodate this at some point in the protocol,
the dealer or players must announce which basis they measure. Results where the bases match are kept, and those that do not match are discarded.
The remaining results should be perfectly correlated and provide a random key,
often referred to as the `sifted key'.
By standard classical security protocols~\cite{GRTZ02},
this can be distilled to a secret key by the dealer and authorised players sacrificing part of the sifted key.

In this way our protocols are protected against a large class of attacks, namely the so-called intercept-resend attacks.
For such attacks we imagine an eavesdropper (traditionally called Eve)
who tries to gain information about the secret key by intercepting the qubits before they reach the players, and either measuring them directly or entangling them to some ancilla `spy'-qubits of her own and then sending them on to the players.
(In fact, this eavesdropper could instead be some unauthorized players.)
At any point in the protocol Eve may measure her entangled ancilla spy-qubits to attempt access to the key.
She will try to do this in a way that is not detectable by the dealer and authorized players.

Measuring in two conjugate bases serves two complementary purposes to counter this particular class of attacks. First the measurements in conjugate bases is  decided randomly,
which protects against any intercept-resend eavesdropping strategy:
if Eve knows which measurement the players will make, she can make the same measurement herself and thus be undetected, but, if she does not know the measurement basis, then any intrusion by her will affect the statistics of subsequent measurements and thus be detectable.
Second, measurement in conjugate bases allows checking of correlations to verify that
the global state indeed corresponds to the correct form~(\ref{eq:conjugate meas}), which proves that
Eve has not entangled the state to ancilla during its transit.

If the measurements are truly random,
we can view them as a kind of tomography of the state shared between the dealer and players, and suppose that all results represent this state faithfully.
In fact these two purposes are not independent and are related to the long history of using entanglement distillation to prove security for key distribution (e.g.~\cite{GRTZ02}). In both cases, the appropriate measurements are the stabilisers of the state.

\begin{figure}
\scalebox{0.45}{\includegraphics*[0.5cm,11cm][21cm,19cm]{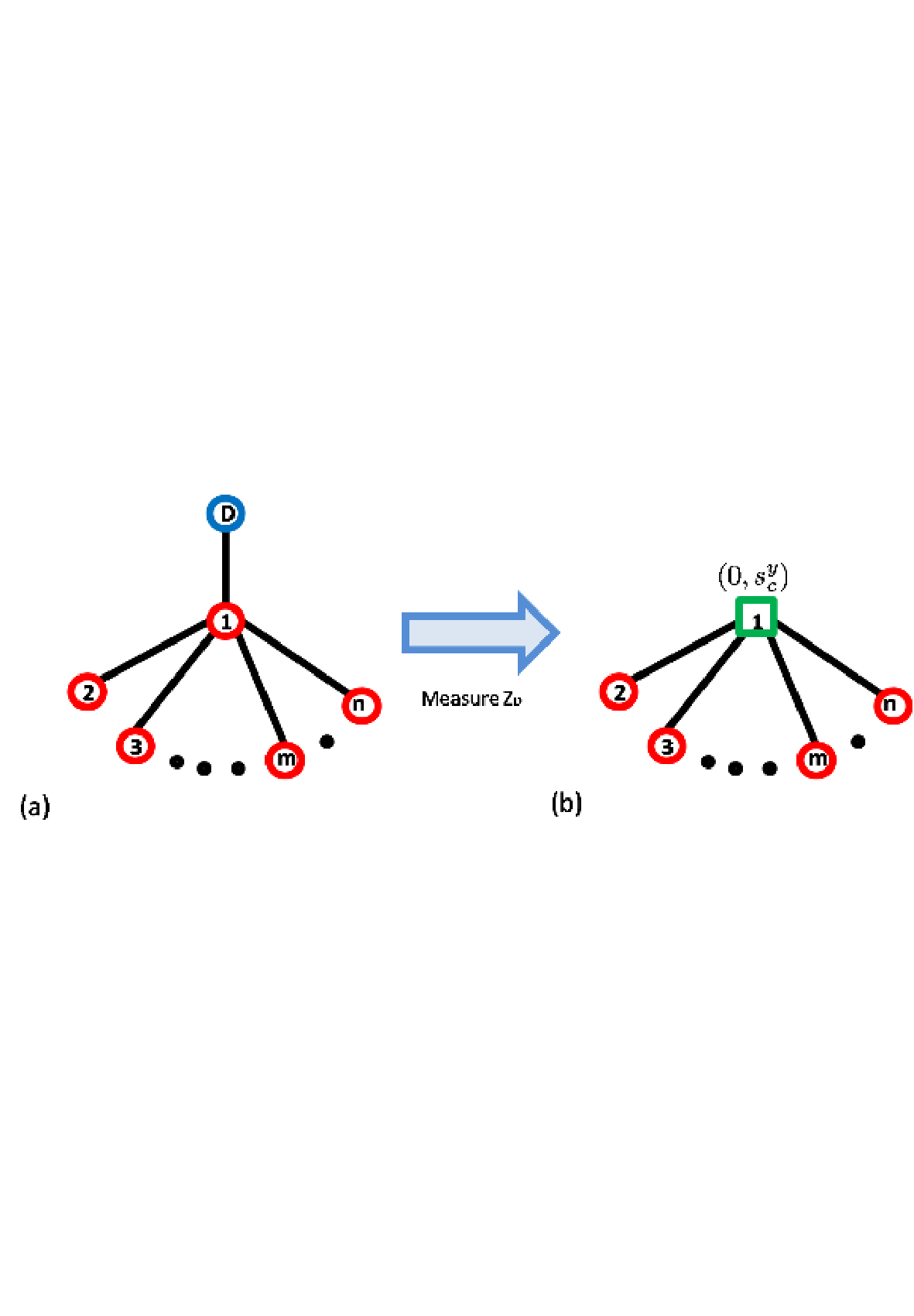}}
\caption{\label{fig:purificationnn}
(Color
 online) {\sf CQ} for (n,n).
The dealer's qubit is attached so as to encode on the first qubit in
the~$(n,n)$ scheme. As Eqs.~(\ref{eq:conjugate meas}) and~(\ref{eq:nGHZkeydistillation}) indicate,
if the authority measures the control
qubit in $X$ the resulting graphs are those of Fig.~\ref{fig:GHZ Dependency}.
In the case of~$Y$ measurements, the graphs transform to the right graph (b) (see
text for explanation).}
\end{figure}

\begin{figure}
\scalebox{0.45}{\includegraphics*[0.5cm,11.5cm][21cm,18.5cm]{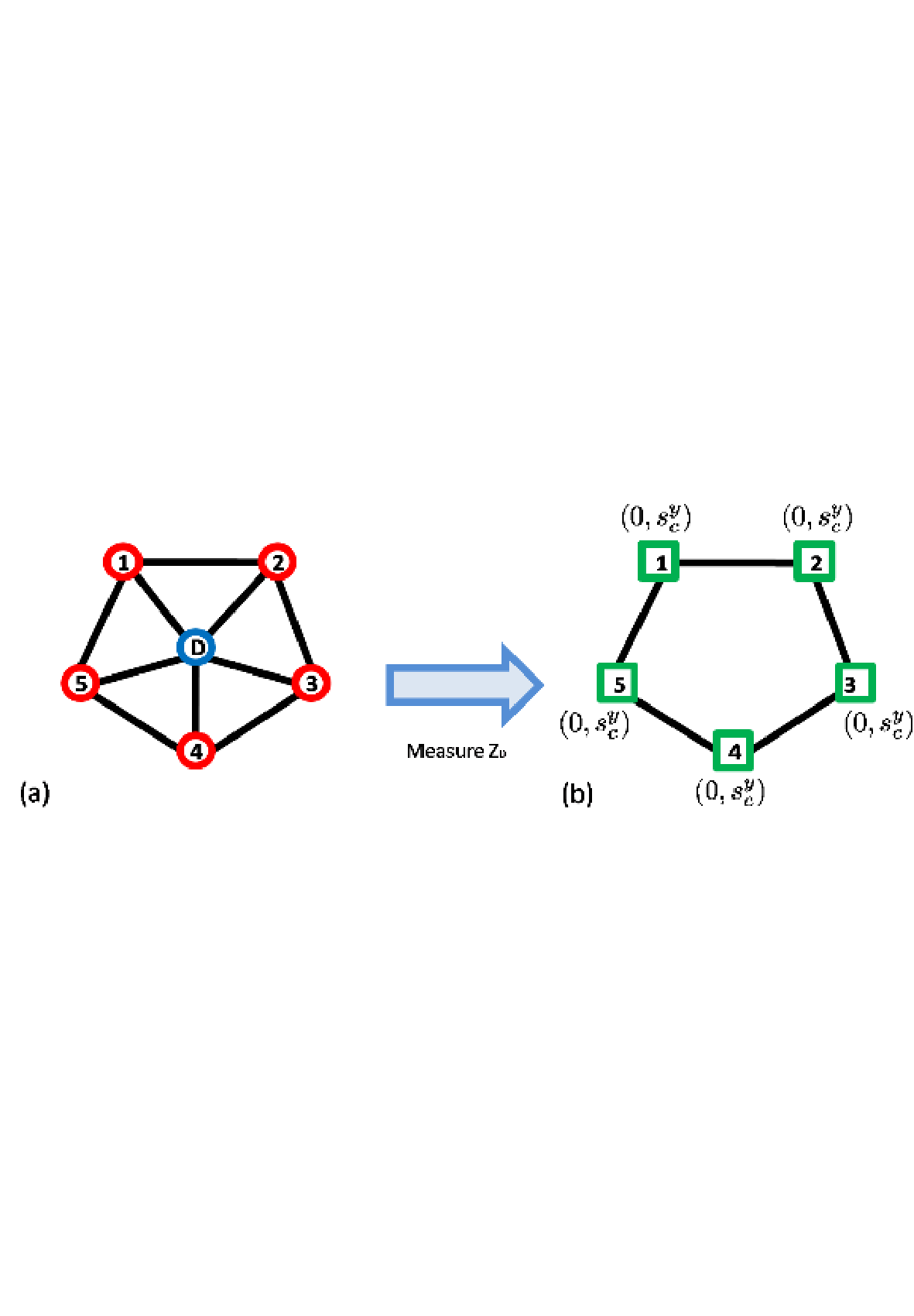}}
 \caption{ \label{fig:purification35}
 (Color
 online) {\sf CQ} for $(3,5)$.
The dealer's qubit is attached so as to encode on all five qubits
correlated for the~$(3,5)$ scheme. As Eq.~(\ref{eq:(3,5) key distilation}) indicates, if the dealer
measures her qubit in $X$ the resulting graph is that of
Fig.~\ref{fig:pentagon}(a).
In the case of $Y$ measurements the graph transform to the
right graph (b).}
\end{figure}

Our protocol runs as follows
\begin{itemize}
\item [{\sf CQ}1] the dealer first prepares the state
$|g_{(0\cdots 0)}\rangle_{1 \ldots n}$ associated to the corresponding CC secret sharing protocol, then entangles extra qubits (known as the dealer's qubits) to each of the qubits upon which the secret is independently distributed (see
Figs.~\ref{fig:purificationnn}a) and~\ref{fig:purification35}a)).
\item [{\sf CQ}2] The dealer distributes to the players their qubits (and keeps the dealer's qubits).
\item [{\sf CQ}3] The dealer and the set of authorized players $\mathsf{V}'$ measure randomly one of a
set of conjugate bases on their qubits and announce they have measured on a public
classical channel.
\item [{\sf CQ}4] The set of players $\mathsf{V}'$ trying to access the secret announce their
measurement basis.
\item [{\sf CQ}5] The dealer announces her measurement basis.
\item [{\sf CQ}6] Classical security protocols follow establishing a secret key between dealer and authorized set of players.
\item [{\sf CQ}7] The secret key is used to send the secret message.
\end{itemize}

Now we see how this works for the~$(n,n)$ case.
The state prepared by the dealer is that of the~$n$GHZM state with an extra vertex attached to player~1's qubit ({\sf CQ}1);
this extra vertex corresponds to a qubit that is retained by the dealer
as shown in Fig.~\ref{fig:purificationnn}(a).
\begin{align}
\label{eq:nGHZkeydistillation}
|\mathcal{G}_{(\vec{0})}\rangle_{D,1,\ldots,n}
	=& \big(|0\rangle_D | g_{(0\cdots 0)}\rangle_{1,\ldots,n} \nonumber \\ &
		+ |1\rangle_D |g_{(10\cdots 0)}\rangle_{1,\ldots,n}\big)/\sqrt{2} \nonumber \\
	=&  (|0'\rangle_D | g_{(0\cdots 0)}'\rangle_{1,\ldots,n} \nonumber \\ &
		-i|1'\rangle_D |g_{(10\cdots 0)}'\rangle_{1,\ldots,n})/\sqrt{2}.
\end{align}
The stabilizers for state~(\ref{eq:nGHZkeydistillation}) are
\begin{align}
\label{eq:nGHZkeydistillation stabilizers}
	K_D^\ocircle =& X_D \otimes Z_1 \otimes \openone_2 \otimes \openone_3 \cdots \otimes \openone_n, \nonumber \\
	K_1^\ocircle =& Z_D \otimes X_1 \otimes Z_2 \otimes Z_3 \cdots \otimes Z_n, \nonumber \\
	K_2^\ocircle =& \openone_D \otimes Z_1 \otimes X_2 \otimes \openone_3 \cdots \otimes \openone_n, \nonumber \\
	K_3^\ocircle =& \openone_D \otimes Z_1 \otimes \openone_2 \otimes X_3 \cdots \otimes \openone_n, \nonumber \\
	\vdots & \nonumber \\
	K_n^\ocircle =& \openone_D \otimes Z_1 \otimes \openone_2 \otimes \openone_3 \cdots \otimes X_n.
\end{align}

The qubit corresponding to vertex~$D$ is retained by the the dealer,
and the others qubits are sent to the players ({\sf CQ}2).
As prescribed by the protocol, the dealer measures in the~$Z$ basis $\{|0\rangle_D, |1\rangle_D\}$ or in the~$Y$ basis $\{|0'\rangle_D, |1'\rangle_D\}$,
and the players perform appropriate measurements to distinguish
$|g_{(0\cdots 0)}\rangle_{1,\ldots,n}$ from $|g_{(10\cdots 0)}\rangle_{1,\ldots,n}$, or $|g'_{(0\cdots 0)}\rangle_{1,\ldots,n}$ from $|g'_{(10\cdots 0)}\rangle_{1\ldots n}$, respectively.
Our arguments from the {\sf CC} case show that discriminating $|g_{(0\cdots 0)}\rangle_{1,\ldots,n}$ from $|g_{(10\cdots 0)}\rangle_{1,\ldots,n}$ cannot be done by a subset of fewer than $n$ players.

Similarly, we can apply properties P1-P4 from Sec.~\ref{subsec:properties} to the conjugate graphs
as shown in Fig.~\ref{fig:purificationnn}(b), to see that the same holds trying to discriminate
$|g'_{(0\cdots 0)}\rangle_{1,\ldots,n}$ from $|g'_{(10\cdots 0)}\rangle_{1,\ldots,n}$.
The players have various choices of how to perform both sets of discriminating measurements. Importantly, the measurements for $g$ will be conjugate to those for $g'$.

We choose the following measurement strategy. The dealer chooses $Z_D$ or $Y_D$ randomly, player one chooses $X_1$ or $Y_1$ randomly and the other players choose $Z_i$ or $Y_i$ randomly. We will see that 50\% of the time their measurements will coincide to allow a shared key. That is, on half of the cases the randomly chosen combination will be equivalent to performing a suitable stabilzer measurement for which the dealer and players will share a perfect key, and also be able to check statistics for security ({\sf CQ}3).
After the announcements of the measurement bases,
the dealer and the players can discard any results where the bases do not match and are left with the sifted key ({\sf CQ}4,5).

To see this, first suppose that the dealer measures in the~$Z_D$ basis.
The players can then obtain a secret key (i.e. discriminate $|g_{(0\cdots 0)}\rangle_{1,\ldots,n}$ from $|g_{(10\cdots 0)}\rangle_{1,\ldots,n}$) by measuring any combination of a product of stabilizers for the subgraph~$g$ as long as they include that of player~1's vertex, which we denote $k_1^\ocircle$
(with lowercase for $k$ indicating a stabilizer for the subgraph rather than the graph).

As we can see by comparing the global state's stabilizers~(\ref{eq:nGHZkeydistillation stabilizers}) and the stabilizers of the subgraph (which can be imagined by extending those of Eq.~(\ref{eq:4GHZMdependencyEG eigen-ops}) to
$n$ parties), including the~$Z_D$ measurement of the dealer, the dealer plus the players are effectively measuring
$K^\ocircle_1$ or any combination of the total graph's stabilizers,
including $K^\ocircle_1$ such as $K^\ocircle_1 \cdot K^\ocircle_i \cdots K^\ocircle_n$.
 Such stabilizer products corresponds to local measurements where player~1 measures $X_1$ and the remaining players measure $Z_i$ or $Y_i$ with an even number of those who measure $Y_i$; or player~1 measures $Y_1$ and the remaining players measure $Z_i$ or $Y_i$ with an odd number measuring $Y_i$.

Similarly, if the dealer measures in the~$Y_D$ basis,
the players can obtain a secret key
by measuring the stabilizers of the conjugate graph depicted in Fig.~\ref{fig:purificationnn}(b),
$k_1^\square$, or any product including this stabilizer:
$k_1^\square \cdot k_i^\ocircle \cdots k_n^\ocircle$ (discriminating
$|g'_{(0\cdots 0)}\rangle_{1,\ldots,n}$ from $|g'_{(10\cdots 0)}\rangle_{1,\ldots,n}$).
Including the measurement by the dealer in the~$Y$ basis,
this set of measurements corresponds to the dealer plus the players measuring stabilizers $K^\ocircle_D \cdot K^\ocircle_1$ or any combination of the graph's stabilizers including $K^\ocircle_D \cdot K^\ocircle_1$, such as $K^\ocircle_D \cdot K^\ocircle_1 \cdot K^\ocircle_i \cdots K^\ocircle_n$. This set of stabilizer measurements corresponds to local measurements where player~1 measures $X_1$ and the remaining players measure $Z_i$ or $Y_i$, as long as there is an odd number of those who measure $Y_i$, or where player~1 measures $Y_1$ and the remaining players measure $Z_i$ or $Y_i$ with an even number of those who measure $Y_i$.

Thus each combination of measurements $X_1$ or $Y_1$ by player one, and~$Z_i$ or $Y_i$ by the other players corresponds to the correct measurement to discriminate the appropriate encoded graph states either for the dealer measuring $Z_D$ or $Y_D$. The dealer chooses randomly so, $50\%$ of the time the measurements will coincide and results be correlated. After checking the bases match they throw away the cases they don't to give a sifted key.

As discussed above, the randomness of the measurement bases ensures that any eavesdropper cannot know which measurement to make, and so any interference she introduces will be noticed by standard security checks (step {\sf CQ}4,5,6) of the protocol.
In this case, the dealer and authorized players should randomly choose a part of the sifted key (for example the dealer picks randomly and announces to all the players), and both the dealer and the authorized players announce their results.

They can then check that the results match those expected. If they do, they are sure there has been no eavesdropper, and that the key is secure. If the correlations do not match, they are forced to throw away the key and start again. As in HBB~\cite{HBB99} the players announce their measurement results ({\sf CQ}4) before the dealer ({\sf CQ}5) to prevent the players from cheating.

That matching results imply security is evident from the fact that the random measurements can be seen as checking the state as mentioned above.
In this case, if measurements are made randomly as described, each choice measures a different set of stabilizers. Doing this asymptotically builds statistics that give the expectation of all the stabilizer operators.
If we obtain the expected results, i.e.\ that the state is a~$+1$ eigenstate of operators $K^\ocircle_D\cdots K^\ocircle_n$ in~(\ref{eq:nGHZkeydistillation stabilizers}), then the state is by definition that of~(\ref{eq:nGHZkeydistillation}) and hence not entangled to any possible eavesdropper so an eavesdropper is denied any information about the key.
We thus have the following proposition.
\begin{proposition} The {\sf CC}~$(n,n)$ scheme can be extended
as above to give secure distribution of shared random keys with the
same secrecy access structure. The secret key can be accessed by LOCC.
\end{proposition}

Let us now see how the~$(3,5)$ can be extended to a secret key distribution protocol. In this case, the state prepared by the dealer is that of Fig.~\ref{fig:purification35}(a):
\begin{align}\label{eq:(3,5) key distilation}
|\mathcal{G}_{(\vec{0})}\rangle_{D12345}
	=& (|0\rangle_D | g_{(00000)}\rangle_{12345} \nonumber \\ &
		+ |1\rangle_D |g_{(11111)}\rangle_{12345})/\sqrt{2} \nonumber \\
	=& (|0'\rangle_D | g_{(00000)}'\rangle_{12345} \nonumber \\ &
		-i|1'\rangle_D |g_{(11111)}'\rangle_{12345})/\sqrt{2}.
\end{align}

The stabilizers of this total state are
\begin{align}
\label{eq:(3,5) key distilation stabilizers}
K_D^\ocircle =& X_D \otimes Z_1 \otimes Z_2 \otimes Z_3 \otimes Z_4 \otimes Z_5, \nonumber \\
K_1^\ocircle =& Z_D \otimes X_1 \otimes Z_2 \otimes \openone_3 \otimes \openone_4 \otimes Z_5, \nonumber \\
K_2^\ocircle =& Z_D \otimes Z_1 \otimes X_2 \otimes Z_3 \otimes \openone_4 \otimes \openone_5, \nonumber \\
K_3^\ocircle =& Z_D \otimes \openone_1 \otimes Z_2 \otimes X_3 \otimes Z_4 \otimes \openone_5, \nonumber \\
K_4^\ocircle =& Z_D \otimes \openone_1 \otimes \openone_2 \otimes Z_3 \otimes X_4 \otimes Z_5, \nonumber \\
K_5^\ocircle =& Z_D \otimes Z_1 \otimes \openone_2 \otimes \openone_3 \otimes Z_4 \otimes X_5.
\end{align}

As before, when the dealer measures in the~$Z$ basis, the fact that the graph $g$ is a~$(3,5)$ direct threshold scheme means that any pair of players cannot reveal the dealer's results, hence cannot access the key;
in contrast any three collaborating players can reveal the key.
As the conjugate $g'$ graph given by the dealer measuring in the~$Y$ basis (see Fig.~\ref{fig:purification35}(b)) is just a reordering of the original ring and replacing $\ocircle\rightarrow\square$ vertices,
it is also a~$(3,5)$ direct thresold scheme, so the same conclusions hold.

Consider first the subset $\{\mathsf{v}_1,\mathsf{v}_2,\mathsf{v}_3\}$:
as the dealer measures $Z_D$,
the players' appropriate measurements correspond to the direct $(3,5)$ protocol of the previous subsection.
That is, player~1 measures $Z_1$, player~2 measures $X_2$, and player~3 measures $Z_3$.
This protocol corresponds to players measuring stabilizer $k_2^\ocircle$ of the induced subgraph (which is the ring of Fig.~\ref{fig:pentagon}),
or the dealer plus the players measuring $K_2^\ocircle$ of the total graph (Fig.~\ref{fig:purification35}(a)).
If the dealer measures $Y_D$, the players' appropriate measurements correspond to the direct $(3,5)$ protocol of the previous section, but changed slightly for the fact that it is the conjugate graph.

By applying the properties of Sec.~\ref{subsec:properties},
we can see this can be done by player~1 measuring $X_1$, player~2 measuring $Y_2$, and player~3 measuring $X_3$.
This corresponds to measuring stabilizer
$k_1^\square \cdot k_2^\square \cdot k_3^\square$ of the induced conjugate graph, which is the ring of Fig.~\ref{fig:purification35}(b), or the dealer plus the players measuring $K_D^\ocircle \cdot K_1^\ocircle \cdot K_2^\ocircle \cdot K_3^\ocircle$ of the total graph
shown in Fig.~\ref{fig:purification35}(a)).

We then have the dealer randomly measuring $Z_D$ or $Y_D$, and the players randomly measuring in $Z_1$, $X_2$, $Z_3$ or $X_1$, $Y_2$, $X_3$.
As before, they communicate to get the sifted key first
(steps~{\sf CQ}4 and~{\sf CQ}5) in the protocol), and then follow the security steps after that.

To observe that the protocol is secure, we study what these measurements imply about the total state of the dealer and the players $\{\mathsf{v}_1,\mathsf{v}_2,\mathsf{v}_3\}$.
In step~{\sf CQ}5 of the protocol, some of the sifted key is sacrificed to verify that the measurement results are those we expected, and if not we throw the key away. If the expectation value of $K_2^\ocircle$ is $+1$, and of $K_D^\ocircle \cdot K_1^\ocircle \cdot K_2^\ocircle \cdot K_3^\ocircle$ is also $+1$, as can be verified by the players, this means that the state of $\{\mathsf{v}_D,\mathsf{v}_1,\mathsf{v}_2,\mathsf{v}_3\}$ is in the subspace of graph states
	$$\left\{|g_{(0000)}\rangle_{D123},|g_{(1100)}\rangle_{D123}, |g_{(1001)}\rangle_{D123},
		|g_{(0101)}\rangle_{D123} \right\},$$
of the subgraph $\{\mathsf{v}_D,\mathsf{v}_1,\mathsf{v}_2,\mathsf{v}_3\}$ given by erasing all other vertices and edges. This also means that there are no correlations between the dealer and any possible eavesdropper. This can be seen as follows.

Given the established subspace, any global state of $\{\mathsf{v}_D,\mathsf{v}_1,\mathsf{v}_2,\mathsf{v}_3\}$ and the environment~$E$
(which may include any eavesdropper) can be written
\begin{align}
\alpha& |0\rangle_{E}|g_{(0000)}\rangle_{D123} + \beta |1\rangle_{E}|g_{(1100)}\rangle_{D123} \nonumber \\
& + \delta |2\rangle_{E}|g_{(1001)}\rangle_{D123} + \gamma |3\rangle_{E}|g_{(0101)}\rangle_{D123}. \nonumber \\
\end{align}
The reduced density matrix of the environment and the dealer can be easily calculated,
using Eq.~(\ref{eq:conjugate meas}),
thereby giving
\begin{equation}
	\rho_{\rm E}\otimes \openone/2_D,
\end{equation}
where $\rho_{\rm E}$ is the reduced density matrix of the environment.
Thus we see that the mutual information between the environment and the dealer is zero;
hence any environment, including a possible eavesdropper, can have no information about the key.

Now consider players $\{\mathsf{v}_1,\mathsf{v}_3,\mathsf{v}_4\}$.
If the dealer measures $Z_D$, the direct $(3,5)$ protocol tells us that the players should measure $X_1$, $Y_3$, $Y_4$, which corresponds to the players measuring the stabilizer product $k_1^\ocircle \cdot k_3^\ocircle \cdot k_4^\ocircle$ of the subgraph (Fig.~\ref{fig:pentagon}), or the dealer plus the players measuring $K_1^\ocircle \cdot K_3^\ocircle \cdot K_4^\ocircle$ of the total graph Fig.~\ref{fig:purification35}(a). If the dealer measures in the~$Y_D$ basis, by considering the discrimination of the conjugate graph states Fig.~\ref{fig:purification35}(b),
the players should measure $Y_1$, $Z_3$, $Z_4$.
Such a measurement protocol corresponds to measuring the stabilizer $k_1^\square$ of the conjugate graph Fig.~\ref{fig:purification35}(b), for the dealer plus the players measuring $K_D^\ocircle \cdot K_1^\ocircle$ of the total graph Fig.~\ref{fig:purification35}(a).

Again the players and the dealer should follow steps {\sf CQ}4 and {\sf CQ}5 to obtain the sifted key
and then compare some results for the security analysis in step {\sf CQ}6.
If the expectation value of $K_1^\ocircle \cdot K_D^\ocircle$ is $+1$, and of $K_1^\ocircle \cdot K_3^\ocircle \cdot K_4^\ocircle$ is $+1$, the state of $\{\mathsf{v}_D,\mathsf{v}_1,\mathsf{v}_3,\mathsf{v}_4\}$ is in the subspace
$$\{|g_{(0000)}\rangle_{D134},|g_{(0011)}\rangle_{D134},
	|g_{(1110)}\rangle_{D134}, |g_{(1101)}\rangle_{D134} \}$$
of the subgraph $\{\mathsf{v}_D,\mathsf{v}_1,\mathsf{v}_3,\mathsf{v}_4\}$ given by erasing all other vertices and edges. Again, any state between the environment and the dealer and players $\{\mathsf{v}_1,\mathsf{v}_3,\mathsf{v}_4\}$,
which has support only in this subspace,
necessarily has no correlations between the dealer and the environment.
Thus again the environment can not get any information about the secret key.

Any set of three players can be covered by these two cases, simply by relabeling the players. We thus have the following proposition.

\begin{proposition} The {\sf CC}~$(3,5)$ scheme can be extended
as above to give secure distribution of shared random keys with the
same secrecy access structure. The secret key can be accessed by LOCC.
\end{proposition}

\begin{remark}
As it stands, this protocol is not resilient to noise. Since noise (for example in the channel between the dealer and players) could interfere with the results of the measurement results in the sifted key, any noise would be seen as a potential eavesdropper and so the sifted key would be discarded.
In fact this is a problem of the original HBB protocol also, and was resolved in~\cite{Chen04} by relating the problem to the distillation of the initial entangled state between the dealer and the players. We expect that the same approach would work for our generalized protocol. In the case of $(n,n)$
 this seems a simple extension.
 However, for the~$(3,5)$ case and possible further graph state bases secret sharing schemes, it would be more subtle and the distillation would be to subspaces with guaranteed security rather than to specific states, reflecting the more complicated security analysis. The large body of work on distilling graph states should prove very useful to this aim (see for example~\cite{HDERNB06} and references therein).
 This is a topic of ongoing work.
 \end{remark}

\subsection{\sf QQ}
\label{subsec:QQ}

We now see how one more extension can be made to allow for the
sharing of quantum secrets.
Here we are addressing the third scenario presented in the introduction {\sf QQ},
namely sending a quantum secret in the possible presence of an eavesdropper. Our approach will be to use the same entangled state as for the key distribution protocol to teleport the secret quantum state across from the dealer to the set of players, where the structure of the graph will again imply the secrecy structure. This teleportation extension is also that used in~\cite{HBB99}, but here we extend it to general graph state protocols and explicitly show it for the~$(n,n)$ and the~$(3,5)$ threshold schemes.

Although we do not present as general a result as those in~\cite{CGL99}, where any $(k,n)$ scheme is shown to be possible, our graph state cases are nevertheless interesting for several reasons.
First, they represent a unification of secret sharing schemes of~\cite{Got00,HBB99,CGL99} via the graph state formalism, second they use only qubits rather than higher dimensional systems as in~\cite{CGL99},
which has advantages in that they are  more easily implementable at present.
Third the schemes here are explicitly written and easy to understand and finally they are written in terms of graph states which are of particular practical interest across quantum information and will be extendible to the embedded protocols of the next section.

Here we should note that because we are sending a quantum secret (that is, a secret quantum state), all players cannot simultaneously obtain the secret if only one is provided, since this would violate no-cloning~\cite{CGL99}.
We say a set of players can access the secret if it sits coherently within the space of their qubits, in a way they know. In this way with access to full quantum channels it can be localised to wherever they choose.

In the case of only classical channels between the players, it may not be possible to localise information at all, and if it is we must choose a site to which they will localise the secret. In each case we discuss where it can be localised if possible by LOCC only.

The protocol operates as follows
\begin{itemize}
\item [{\sf QQ}1]
The dealer prepares a secrecy graph state as in the {\sf CQ} key
distribution protocol.
\item [{\sf QQ}2]
The dealer performs a Bell measurement on the qubit $D$ and her input
secret qubit $|S\rangle_\text{in}$, and performs the appropriate correction operation.
\item [{\sf QQ}3]
The dealer distributes to the players their qubits.
\item [{\sf QQ}4]
Authorised sets of players access the secret by making prescribed measurements and corrections to focus down the information to one of the authorized parties. This may require global operations by authorized players (requiring quantum channels).
\end{itemize}

Let us see how this works. After step~{\sf QQ}1 the dealer possesses the encoded graph state $|\mathcal{G}_{(\vec{0})}\rangle_{D,1,\ldots,n}$ and the input quantum secret $|S\rangle_{\rm in} = \alpha |0\rangle_{\rm in} + \beta |1\rangle_{\rm in}$. We can see what the resultant state of the players will be after step~{\sf QQ}2 by rewritting the state in the Bell basis
\begin{align}
	|B_{00}\rangle:=&(|00\rangle+|11\rangle)/\sqrt{2},	\nonumber	\\
	|B_{01}\rangle:=&(|00\rangle-|11\rangle)/\sqrt{2},	\nonumber	\\
	|B_{10}\rangle:=&(|01\rangle+|10\rangle)/\sqrt{2},	\nonumber	\\
	|B_{11}\rangle:=&(|01\rangle-|10\rangle)/\sqrt{2},
\end{align}
as follows.
\begin{widetext}
\begin{align}
|S\rangle_{\rm in} |\mathcal{G}_{(\vec{0})}\rangle_{D,1,\ldots,n}
=& 1/4 (|B_{00}\rangle_{\text{in},D} (\alpha|g_{(0\cdots 0)}\rangle_{1,\ldots,n}
	+ \beta|g_{(\underbrace{1\cdots 1}_{\in N_D}0\cdots 0)}\rangle_{1,\ldots,n}
		\nonumber	\\	&
	+ |B_{01}\rangle_{\text{in},D} (\alpha|g_{(0\cdots 0)}\rangle_{1,\ldots,n}
	- \beta|g_{(\underbrace{1\cdots 1}_{\in N_D}0\cdots 0)}\rangle_{1,\ldots,n} \nonumber \\ &
	+ |B_{10}\rangle_{\text{in},D} (\alpha|g_{(\underbrace{1\cdots 1}_{\in N_D}0\cdots 0)}\rangle_{1,\ldots,n} +
	\beta|g_{(0\cdots 0)}\rangle_{1,\ldots,n}
		\nonumber	\\	&
	+ |B_{11}\rangle_{\text{in},D} (\alpha|g_{(\underbrace{1\cdots 1}_{\in N_D}0\cdots 0)}\rangle_{1,\ldots,n} -
	\beta|g_{(0\cdots 0)}\rangle_{1,\ldots,n}).
\end{align}
\end{widetext}
This can be seen using Eq.~(\ref{eq:conjugate meas}).
After obtaining result $B_{ij}$ corresponding to Bell state $|B_{ij}\rangle$, the correction operator
\begin{equation}
	U(i,j)_{1,\ldots,n}:= \otimes_{a\in N_c}Z_i^i \cdot K_{a\in N_c}^{\ocircle j},
\end{equation}
brings the resultant state to the desired encoded secret state
\begin{equation}
\label{eq:qsecret}
	|S_g\rangle = \alpha|g_{(0\cdots 0)}\rangle_{1,\ldots,n} + \beta|g_{(\underbrace{1\cdots 1}_{\in N_C}0\cdots 0)}\rangle_{1,\ldots,n}.
\end{equation}

In fact, Eq.~(\ref{eq:qsecret}) is analogous to the direct encoding approach of
\cite{CGL99,Got00}, where a state is encoded directly onto a stabilizer space, or codespace. Again~\cite{CGL99,Got00} is more general than ours in that it covers all $(k,n)$, but again our scheme has the advantage of being set on qubits and explicilty and simply written down (where as in~\cite{CGL99,Got00} it is done with high dimensional stabilizer codes via error correction codes and is not readily understandable in terms of states).

To determine the access structure,
that is to see if a set of players can access the secret, we study the reduced density matrix. From Eq.~(\ref{eq:qsecret}), it is clear that same adversary structure (i.e. the sets of players who cannot access the secret) as in the {\sf CQ} case  cannot access the secret perfectly here. For example, by setting either $\alpha=0$ or $\beta=0$ in Eq.~(\ref{eq:qsecret}),
we can apply directly the dependency properties (c.f.\ Sec.~\ref{subsec:properties}). However, it is possible that some some information (though not perfect) may be obtained because of the cross-over terms in the reduced density matrix. In addition, the access of information can no longer be checked using previous methods, as it is quantum not classical and different rules can apply (such as no-cloning). In other words, we also have to check whether the quantum secret state is coherently placed within the reduced density matrix, and if it is, whether it can be localized by LOCC.

This requirement illustrates the difference between sharing classical and quantum secrets as discussed in detail
by Gottesman~\cite{Got00}. We will see that, for the~$(n,n)$ threshold schemes, the access structure
remains the same as for the {\sf CQ} case, although the adversary structure can be significantly different (indeed any individual player can access some information about the secret). For the~$(3,5)$ threshold scheme both the access and adversary structure remain the same. For a more general approach to the question of access, links to the flow conditions for determinism in measurement-based quantum computation~\cite{Dan06} are quite useful, as we see in our example below. We leave this to later work and for now concentrate on these two examples. 

In the~$(n,n)$ case the initial state prepared by the dealer is the same as that for the key distilation protocol, shown in Fig.~\ref{fig:purificationnn}(a).
After the teleportation steps~{\sf QQ}2 and~{\sf QQ}3,
the players share the state
\begin{equation}
\label{eq:qsecretnn}
	|S_{g}\rangle_{1,\ldots,n}
		= \alpha |g_{(0\cdots 0)}\rangle_{1,\ldots,n} + \beta|g_{(10\cdots 0)}\rangle_{1,\ldots,n},
\end{equation}
encoding the quantum secret.
The fact that any set of $n~-~1$ players cannot access the secret perfectly follows from the dependency property from Sec.~\ref{subsec:properties} applied to the graph $g$, which, as in the case of the key distribution protocol, means the states $|g_{(\vec{0})}\rangle$ and~$|g_{(10\cdots 0)}\rangle$ appear locally identical to any such set, and thus implies they cannot read the secret perfectly. However, the cross over terms in the reduced density matrix means that any subset of players (including a single player) can access some information. For a set of players $A$, the reduced density matrix is of the form
\begin{equation}
\rho_A(S) = \frac{|\alpha +\beta|^2}{2} |\bar{0}\rangle \langle\bar{ 0}| + \frac{|\alpha -\beta|^2}{2}|\bar{1}\rangle\langle\bar{1}|,
\end{equation}
where $|\bar{0}\rangle,|\bar{1}\rangle$ are computational basis states over $A$ (up to local Hadamards).
Thus, for secret states $|S\rangle_\text{in}$ lying on the $Z-Y$ plane of the Bloch sphere, all reduced states look the same;
hence no information can be gained in this case. However, for other states, some information can be obtained. For example, any single player could distinguish perfectly the two secret states
$|S\rangle_\text{in}=(|0\rangle \pm |1\rangle)/\sqrt{2}$.

On the other hand by teleportation the secret state is coherently shared by all players, hence if acting all together, the players can access the secret perfectly, and can certainly localise it to any chosen player by using quantum channels. If restricted to LOCC, they can in fact, also localise the secret onto any player. First we consider localising it to any player other than player~1.
We can see this explicitly if rewrite the corresponding state~(\ref{eq:qsecret}) using~(\ref{eq:conjugate meas}) in the following way
\begin{align}
	|S_{g}\rangle_{1,\ldots,n}
		=& \alpha |g_{(0\cdots 0)}\rangle_{1,\ldots,n}
			+ \beta|g_{(10\cdots 0)}\rangle_{1,\ldots,n} \nonumber \\
		=& |g_{(0\cdots 0)}\rangle_{1,\ldots,n / i} (\alpha |0\rangle_i +\beta |1\rangle_i) \\ \nonumber
			& + |g_{(10\cdots 0)}\rangle_{1,\ldots,n/i} (\beta |0\rangle_i +\alpha |1\rangle_i);
\end{align}
thus if player~1 measures $X$ and the remaining players except player $i$ measure $Z$, they can push the quantum state onto player $i$'s qubit. To push the secret state onto player~1, it is enough for the remaining players to measure in the~$Z$ basis (and of course communicate their result to player~1),
as can be seen by rewriting the state as
\begin{widetext}
\begin{align}
	|S_{g}\rangle_{1,\ldots,n}
		=& \alpha |g_{(0\cdots 0)}\rangle_{1,\ldots,n}
			+ \beta|g_{(10\cdots 0)}\rangle_{1,\ldots,n} \nonumber \\
		=& \sum_{m_2 \oplus\ldots  \oplus m_n=0}|m_2 \ldots m_n\rangle_{2,\ldots,n} (\alpha |+\rangle_1
			+\beta |-\rangle_1)
		+ \sum_{m_2\oplus \ldots  \oplus m_n=1}|m_2 \ldots m_n\rangle_{2,\ldots,n}  (\alpha |-\rangle_1
			+\beta |+\rangle_1),
\end{align}
\end{widetext}
where $m_i$ is a bit and $|m_2 \ldots m_n\rangle_{2,\ldots,n}$ is a product of Pauli $Z$ eigenstates.

We can see this another way, which may give hints of a possible way to check accessibility in more general cases. This starts from noting that Eq.~(\ref{eq:qsecret}) as having arisen from another scheme, not teleportation, analogous to the error correcting schemes in~\cite{Schlingemann01}, uniting these approaches. To do this the dealer entangles the secret state $|S\rangle$ onto vertex $1$ of the~$n$GHZM state directly and measures it in the~$Z$ basis. By making the appropriate correction (based on the measurement outcome), the resultant state is that of~(\ref{eq:qsecret}).

Incidently, we notice that our five party scheme is the same graph as the 5-qubit code in
\cite{Schlingemann01}, which is perhaps not so surprising, since sharing a threshold scheme for quantum secrets is
necessarily an error correcting scheme (though not the other way round) as pointed out in~\cite{CGL99,Got00}. Indeed this example appears in~\cite{CGL99,Got00} (although established in a different way).

Viewing Eq.~(\ref{eq:qsecretnn}) as having arisen in this way, we can imagine the state of the secret qubit entangled to the~$n$GHZM state as the start of a measurement based computation with the secret qubit as the input. We then choose any one player as an output and application of the Flow conditions~\cite{Dan06,BKMP07} give an explicit way to direct the secret onto any one of the players' qubits by LOCC.

We thus have the following proposition.

\begin{proposition}  \label{prop:(n,n) QQ} The {\sf CQ}~$(n,n)$ scheme can be extended
as above to give quantum secret sharing schemes with a weaker access structure: any fewer than $n$ players cannot access the secret perfectly, although they may access some information, and all $n$ players working together can access the secret perfectly. The secret can be localised onto any one of the authorized players by LOCC.
\end{proposition}
We note that the impossibility of achieving the $(n,n)$ threshold scheme follows directly from Ref.~\cite{CGL99} as this is a pure state scheme in their terms. However, it has recently been shown to be possible, using a mixed state scheme by Broadbent et al~\cite{BCT09}, which can be understood as a classical mixing over the scheme above, so that any sets of players fewer than $n$ see no information.

In the~$(3,5)$ threshold case, we will see that the threshold access structure carries through.
The dealer starts with the state of Fig.~\ref{fig:purification35}(a). After teleportation steps ii) and iii) in the protocol, the players have the state
\begin{equation}
|S_g\rangle_{12345} = \alpha |g_{(00000)}\rangle_{12345}+ \beta|g_{(11111)}\rangle_{12345}.
\end{equation}
 It can easily be checked that the reduced density matrices of any sets of fewer than three players are identical for all $\alpha$, $\beta$, hence they cannot access any information about the secret qubit.

The ability for three users to access the secret qubit can be seen to follow from the
fact that it is also a two qubit erasure error correcting scheme~\cite{Schlingemann01}. In general this
is not LOCC. We can see how this works explicitly writing the state,
\begin{align}
|S_g\rangle_{12345} =& \alpha |g_{(00000)}\rangle_{12345}+ \beta|g_{(11111)}\rangle_{12345} \nonumber \\
	=& 1/4 ( |B_{00}\rangle_{13}(\alpha|+\rangle_2 + \beta|-\rangle_2)|B_{01}\rangle_{45}
			\nonumber	\\	&
	+ |B_{01}\rangle_{13}(\alpha|+\rangle_2 - \beta|-\rangle_2)|B_{10}\rangle_{45}
			\nonumber	\\	&
	+ |B_{10}\rangle_{13}(\alpha|-\rangle_2 - \beta|+\rangle_2)|B_{00}\rangle_{45}
			\nonumber	\\	&
	+ |B_{11}\rangle_{13}(\alpha|-\rangle_2 + \beta|+\rangle_2)|B_{11}\rangle_{45}),
\end{align}
where $|\pm\rangle = 1/\sqrt{2}(|0\rangle \pm |1\rangle)$. It is clear that if either pair, $\{\mathsf{v}_1,\mathsf{v}_3\}$ or $\{\mathsf{v}_4,\mathsf{v}_5\}$ measure in the Bell basis, they can localize the quantum secret onto player~2's qubit by telling her the result. After the Bell measurements, player~2's qubit will be projected into one of four possible states as seen above, which can be brought to $|S\rangle_2$ by the appropriate correction operation depending on the result (this will consist of a local Hadamard and possibly local $X$ and~$Z$ flips). Similarly any set of three players can localize the secret onto one of their number by two getting together and performing a Bell measure. This leads to the following proposition.

\begin{proposition} The {\sf CQ}~$(3,5)$ scheme can be extended
as above to give quantum secret sharing schemes with the same
secrecy access structure. The secret can be localised onto one of the players in an authorized set using quantum channels.
\end{proposition}

In~\cite{CGL99} it is shown that there can exist no (3,4) QSS scheme
using pure states - hence it is clear there is no way to extend the~$(3,4)$ scheme in this way.
This intriguingly coincides with the fact that we cannot see how to do a {\sf CQ} for $(3,4)$ and may indicate a more fundamental connection.

\section{Embedded protocols}
\label{sec: embedded}

We now present an example of an embedded protocol. That is, we embed the~$(n,n)$ protocols of the previous section into a larger graph, taking advantage of the fact that graph states are extremely useful for various quantum information tasks. The setting we imagine is where a quantum network is envisaged comprising of graph states, and we would like to do secret sharing on some sub-network of the total network. This may occur as a protocol in its own right or as part of a more involved protocol, possibly involving error correction and or MBQC.

\begin{figure}
\scalebox{0.43}{\includegraphics*[0cm,9cm][21cm,20.5cm]{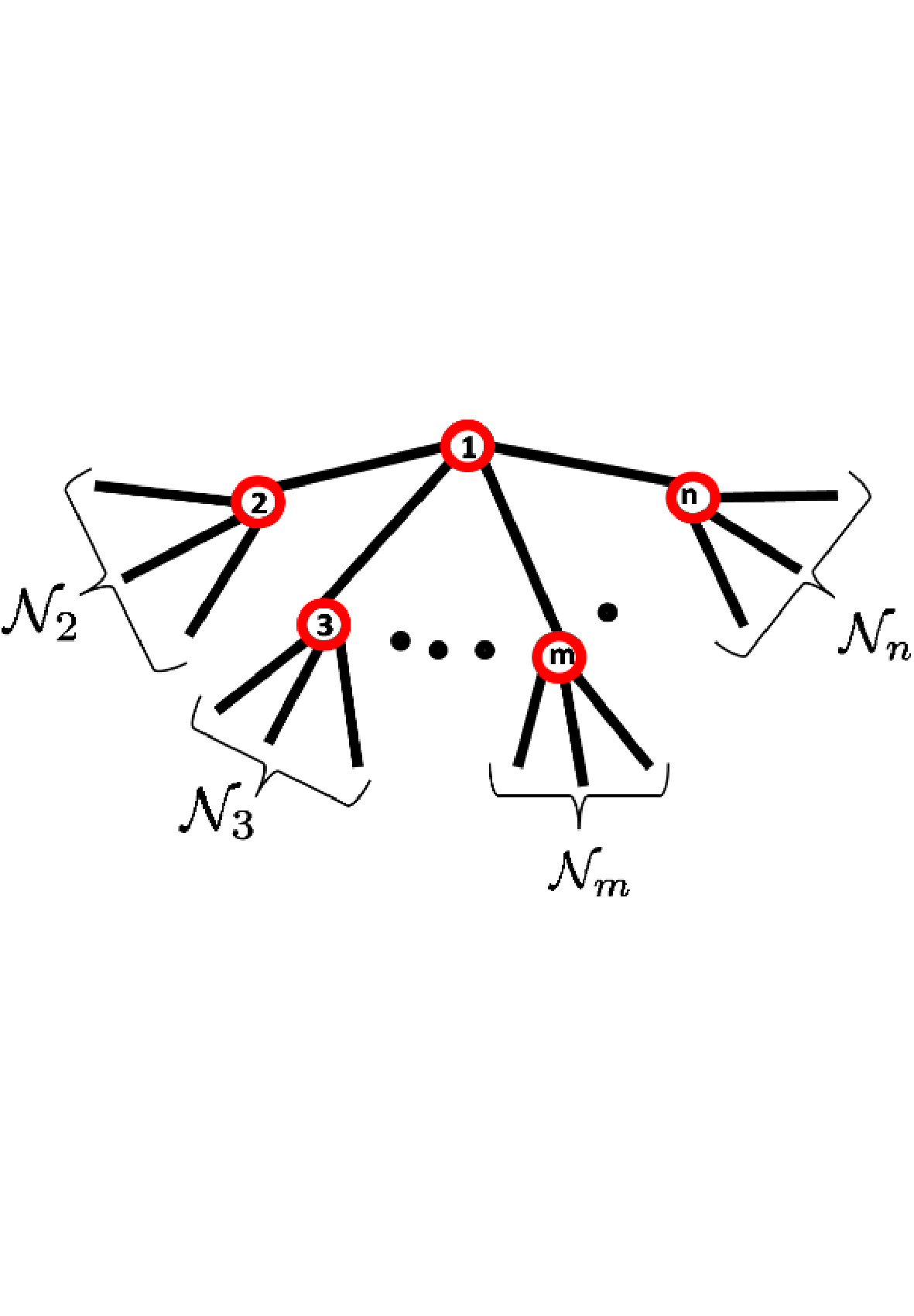}}
  \caption{\label{fig:Tree}
    (Color
 online) $n$-GHZ tree state. $\mathcal{N}_i$ represents the neighbours of qubit $\mathsf{v}_i$ (minus $\mathsf{v}_1$), and they are allowed to overlap. This state gives a~$(n,n)$ secret sharing schemes for qubits $1$ to $n$
    (see propositions \ref{prop:Tree direct}, \ref{prop:Tree key distribution} and \ref{prop:Tree QSS}).}
\end{figure}

We do this via a simple extension of a~$n$-party GHZ
states, which we call a {\it $n$-GHZ tree} state, depicted in Fig.~\ref{fig:Tree}.
\begin{definition}
A~$n$-GHZ tree state is any state where there is an $n$GHZM state
embedded inside another graph, attached by any of the vertices
except $\mathsf{v}_1$.
\end{definition}

We now imagine a situation where the dealer hands out this state to be shared by the players $\{\mathsf{v}_1,\ldots,\mathsf{v}_n\}$ and the rest of the network, in this case the neighours $\mathcal{N}_2,\ldots,\mathcal{N}_n$ (which may overlap). We then ask, can we still use this state to do the secret sharing protocols of the last three sections. We find that we can, starting with the following proposition.

\begin{proposition} \label{prop:Tree direct}
For all $n$-GHZ trees of (see Fig.~\ref{fig:Tree}),
the bit $S=l_{12}$ forms an embedded $(n,n)$
direct {\sf CC} classical secret sharing scheme (within the subgroup of players $\{\mathsf{v}_1,\ldots,\mathsf{v}_n\}$). This is true
independent of the form of all possible sets of neighbours $\mathcal{N}_i$
(including overlapping).
\end{proposition}

The protocol would run as in the case of Sec.~\ref{subsec:direct},
with the secret being encoded onto the bit $S=l_{12}$ on the~$n$-GHZ tree. The validity of the proposition above can be seen since $l_{12}$ can, and only can, be read by all $n$ of the subset of $ \{ 1,\ldots,n \} $ players together.
This requirement is implied by simple application of the dependency and access properties, which are unaffected by the addition of neighbours in this case.
However, there is a subtlety introduced if the neighbours of one of the players may help.

For example, if a neighbour of player~2, say $\mathsf{v}_j\in \mathcal{N}_2$, who, for simplicity's sake we say only neighbours $2$ and no one else, the secret $S=l_{12}$ can be read by the players
$$\{\mathsf{v}_1,\mathsf{v}_3,\mathsf{v}_4,..,\mathsf{v}_n,\mathsf{v}_j\}$$
since they can measure $K_1^\ocircle \cdot K^\ocircle_{j \in \mathcal{N}_2}$ (as it is nontrivial only on those players' systems). Thus they can find $l_{12}\oplus l_{j2} = S$ without player~2. However, when not allowing for the help of the neighbours, and only allowing players in the set $\{\mathsf{v}_1,\ldots,\mathsf{v}_n\}$ to act together, all $n$ must co-operate to find the secret, hence it acts as an $(n,n)$ threshold scheme.

This would not work for the square $(3,4)$ or $(3,5)$ schemes since the secret is shared amongst all the players, so attaching neighbours to any would spread it out and hence it could not be found without measuring the neighbours also, by P2.

This embedded structure can also be extended to key distribution and quantum secret sharing. As in the~$(n,n)$ key distribution case the total state is made by taking the {\sf CC} case and attaching a vertex to player~1, which is kept by the dealer. In the case of the~$n$-GHZ tree state, the total state of dealer, players and neighbours is given by
\begin{align}
\label{eq:embedded key distilation}
|\mathcal{G}_{(\vec{0})}\rangle&_{D,1,\ldots,n,N_2..N_n} \nonumber \\
	=&( |0\rangle_D |g_{(0\cdots 0)}\rangle_{1,\ldots,n\mathcal{N}_2..\mathcal{N}_n}  \nonumber \\ &
	+ |1\rangle_D |g_{(10\cdots 0)}\rangle_{1,\ldots,n \mathcal{N}_2..\mathcal{N}_n}) \nonumber \\
	=& \frac{e^{i\pi/4}}{\sqrt{2}}( |0^{'}\rangle_D |g_{(0\cdots 0)}^{'}\rangle_{1,\ldots,n\mathcal{N}_2..\mathcal{N}_n} \nonumber \\
&
	-i |1^{'}\rangle_D |g_{(10\cdots 0)}^{'}\rangle_{1,\ldots,n\mathcal{N}_2..\mathcal{N}_n}).
\end{align}

The protocol would then run as in Sec.~\ref{subsec:CQ}, using this state. As explained above, if players $\{\mathsf{v}_1,\ldots,\mathsf{v}_n\}$ work together,
they can discriminate
$$|g_{(0\cdots 0)}^{(')}\rangle_{1,\ldots,n\mathcal{N}_2..\mathcal{N}_n}$$ and
$$|g_{(10\cdots 0)}^{(')}\rangle_{1,\ldots,n\mathcal{N}_2..\mathcal{N}_n}$$
because they can measure $k_1^{\ocircle,\square}$ as they only have support on $\{\mathsf{v}_1,\ldots,\mathsf{v}_n\}$. Hence they can access the key.
With the dealer and the players these measurements correspond to measuring stabilizers $K_1^\ocircle$ and~$K_D^\ocircle \cdot K_1^\ocircle$ of the total graph (that of Fig.~\ref{fig:Tree} with the dealer attached to vertex 1). Again, from the dependency arguments P1 any subset of $\{\mathsf{v}_1,\ldots,\mathsf{v}_n\}$ cannot, though again, the same subtleties hold regarding the help of the neighbours.

With regards to security against an eavesdropper, unfortunately we cannot check all stabilizers in here as we did in the~$(n,n)$ case, since the remaining stabilizers have support outside qubits $\{1,\ldots,n\}$. However, we can check that it remains in a secure subspace, as for the $(3,5)$ case in Sec.~\ref{subsec:CQ}.

Let $h$ denote the subgraph induced by removing all neighbouring sets $\{\mathcal{N}_i\}$ from Fig.~\ref{fig:Tree} plus a dealer vertex attached to $\mathsf{v}_1$ (giving the graph as in Fig.~\ref{fig:purificationnn}(a)), and $|h_{\vec{\ell}_{\star 2}}\rangle$ the associated encoded graph state. If the expectation values of $K_1^\ocircle$ and~$K_D^\ocircle \cdot K_1^\ocircle$ are both $+1$ (as can be checked be sacrificing part of the sifted key), then we can be sure that the state of the dealer and players $\{\mathsf{v}_1\ldots \mathsf{v}_n\}$  are in the subspace
$$\left\{|h_{00\vec{\eta}_{\star2}^{(2,\ldots,n)}}\rangle_{D 1,\ldots,n}\right\}.$$
That is, in the subspace such that the first two entries of the bit string are zero, and the remaining $n-1$ elements are free,
which we denote with bit string $$\vec{\eta}_{\star2}^{(2,\ldots,n)}$$
(i.e., the subspace spanned by all such bit strings).

If we now call $p$ the graph after also removing the dealer's qubit (i.e. the $nGHZM$ graph), the complete state of the dealer, players $\{\mathsf{v}_1\ldots \mathsf{v}_n\}$, and any environment $E$
(including the neighbours and possible eavesdropper) can therefore be written as
\begin{align}
|\Psi\rangle_{ED1,\ldots,n}=&
	2^{-1/2} \sum_{\vec{\eta}_{\star2}^{(2,\ldots,n)}} \alpha_{\vec{\eta}_{\star2}^{(2,\ldots,n)}}
	\big(|0\rangle_D|
	p_{0\vec{\eta}_{\star2}^{(2,\ldots,n)}}\rangle_{1,\ldots,n}
		\nonumber \\ &
	+ |1\rangle_D|p_{1\vec{\eta}_{\star2}^{(2,\ldots,n)}}
	\rangle_1,\ldots,n\big)|\vec{\eta}_{\star2}^{(2,\ldots,n)}\rangle_E,
\end{align}
where the summation is over the bit string $\vec{\eta}_{\star2}^{(2,\ldots,n)}$.
Tracing out all the players, we have that the state of the dealer and the environment is
$\rho_{\rm E}\otimes \openone/2_D$;
thus we see that the mutual information between the environment and the dealer is zero.

Hence, on the condition that the value of $K_1^\ocircle$ and~$K_D^\ocircle \cdot K_1^\ocircle$ are both found to be $+1$, any environment, including a possible eavesdropper, can have no information about the key. If the measurement results do not satisfy this condition,
they should be discarded as in Sec.~\ref{subsec:CQ}. Thus, we thus have the following proposition.

\begin{proposition}
\label{prop:Tree key distribution}
For all $n$-GHZ trees of $n$ qubits (see Fig.~\ref{fig:Tree}),
attaching the dealer's qubit to vertex $\mathsf{v}_1$ forms an embedded $(n,n)$
key distribution {\sf QC} secret sharing scheme (within the subgroup of players $\{\mathsf{v}_1,\ldots,\mathsf{v}_n\}$). This is true
independent of the form of all possible sets of neighbours $\mathcal{N}_i$
(including overlapping).
\end{proposition}

We can also see how to extend this embedded protocol to sharing a quantum secret. As in Sec.~\ref{subsec:QQ}, the dealer uses the state of the key distillation protocol, in this case~(\ref{eq:embedded key distilation}), to teleport the secret qubit onto the player's qubits
giving the encoded state dealt to the players and their neighbours
\begin{widetext}
\begin{align}
|S_{g}\rangle_{1,\ldots,n, \mathcal{N}_2..\mathcal{N}_n}
	=&  \alpha |g_{(0\cdots 0)}\rangle_{1,\ldots,n, \mathcal{N}_2,\ldots,n_n}
		+ \beta |g_{(10\cdots 0)}\rangle_{1,\ldots,n, \mathcal{N}_2,\ldots,n_n}	\nonumber \\
	=& \sum_{m_2 \oplus \cdots  \oplus m_n=0}|m_2 \ldots m_n\rangle_{2,\ldots,n}
		|\mathcal{N}_{\vec{m}_{\star 2}}\rangle_{\mathcal{N}_2 \ldots \mathcal{N}_n}
		\left(\alpha |+\rangle_1 +\beta |-\rangle_1\right)	\nonumber	\\
	& + \sum_{m_2 \oplus \cdots  \oplus m_n=1}|m_2 \ldots m_n\rangle_{2,\ldots,n} |\mathcal{N}_{\vec{m}_{\star 2}}\rangle_{\mathcal{N}_2 \ldots \mathcal{N}_n} \left(\alpha |-\rangle_1 +\beta |+\rangle_1\right),
\end{align}
\end{widetext}
with $\mathcal{N}$ the subgraph of the set $\{\mathcal{N}_2 \ldots \mathcal{N}_n\}$,
$|\mathcal{N}_{\vec{m}_{\star 2}}\rangle_{\mathcal{N}_2 \ldots \mathcal{N}_n}$ the associated encoded graph state, and bit string $\vec{m}_{\star2}$ is a function of bit values $\{m_2, \ldots, m_n\}$.

From this it is clear that if players $\{\mathsf{v}_2,\ldots,\mathsf{v}_n\}$ measure in the~$Z$ basis, they project the state of player~1's qubit onto either $\alpha |+\rangle_1 +\beta |-\rangle_1$ or $\alpha |-\rangle_1 +\beta |+\rangle_1$ if the total parity is odd or even respectively. These can be taken to the original secret state $|S\rangle_1$ by the action of correction operator $U^\text{even}_1= H_1$ for even parity, and~$U^\text{odd}_1=H_1 Z_1$ for odd parity.  Again we have a weaker version of the condition for the adversary structure. The dependency rule P1 means that any less than the complete set from the set of players $\mathsf{v}_1,\ldots,\mathsf{v}_n$ cannot access the secret perfectly, but any single player can access some information. Thus we have the following proposition illustrated in Fig.~\ref{fig:Tree}.

\begin{proposition} \label{prop:Tree QSS}
For all $n$-GHZ trees of $n$ qubits
attaching the dealer's qubit to vertex $\mathsf{v}_1$ forms an embedded weak $(n,n)$
{\sf QQ} quantum secret sharing scheme with the same adversary and access structures as Proposition \ref{prop:(n,n) QQ} (within the subgroup of players $\{\mathsf{v}_1,\ldots,\mathsf{v}_n\}$). This is true
independent of the form of all possible sets of neighbours $\mathcal{N}_i$
(including overlapping).
\end{proposition}

The {\sf QQ} is perhaps the most useful of the embedded protocols as it could readily be seen as part of an extended network, for example, the input to our protocol could occur as an output of some previous MBQC, and the output of our protocol could be the input to a subsequent computation.

\section{Conclusions}
We have seen that graph states are a good resource giving a unified approach to secret sharing of both quantum and classical secrets, in the presence of an eavesdropper. The protocols presented can be clearly understood from simple properties of classical information on graph states we have showed. In particular this gives a new way to generate protocols of addressing the second situation, that of sharing a quantum secret in the presence of eavesdropper, beyond $(n,n)$ for the first time. Explicilty we have presented the case of a~$(3,5)$ secret sharing protocol for such a setting. Further we have introduced so-called embedded secret sharing protocols as a first step to integrated quantum networks.

A natural question to ask next is how far can these ideas be pushed and which $(k,n)$ can be covered. This then becomes a graph theoretic relabeling
problem - where the labels are the secrets (more precisely the encoding of the secrets), and their access given by our rules. We expect that this way will result in many cases. However, we know that directly not all $(k,n)$ cases can be solved. We can see for example that it is impossible to do a~$(3,2)$ direct secret sharing scheme with a graph state using our protocol. This can be seen easily, firstly by noticing that demanding single individuals get no information means all three must be entangled. Then we have two options, either the~$3$-GHZM, or the fully connected graph. In both cases it is easy to see that there is no way to choose an encoding of secret~$S$ onto the~$\ell_{i2}$ such that all three pairings depend on it simultaneously (thus, they cannot all access it). The application of the graphical properties presented to this problem is an ongoing area of investigation.

We can also ask is it possible to extend the protocols in this work somehow to cover these cases which currently cause difficulty.
One such extension may be possible via the connection between the sharing of quantum and classical secrets we have seen in our protocols. In~\cite{CGL99} solutions to the general $(k,n)$ sharing of a quantum secret were provided via connections made to error correction.

Although we arrived at our {\sf QQ} from a different approach,not related to error correction, they can be thought of as stabilizer codes, so relating to those of~\cite{CGL99}, since all stabilizer codes can be thought of as graph codes~\cite{Schlingemann01}. A benefit of our approach is that we have been able to use intuitive graphical ideas to prove which players can access which information. Perhaps extending to higher dimensional graph states, we would be able to recapture the results of~\cite{CGL99} for the most general case seems likely to be possible for sharing quantum secrets. The next step would be to translate this to the key distribution protocol, which would be possible via the connections shown here.

As we have mentioned, there is also the outstanding questions of security for these protocols.
Here we have seen security against certain classes of attacks, as in HBB~\cite{HBB99}. The question of security for secret sharing is a large problem of current interest~\cite{Demkowicz-Dobranski08,MatsumotoR07,SS05}. Another problem raised was that of noisy channels as it stands our protocol is not resilient to noise. This was also a problem for HBB, and it seems likely that the same approach used in~\cite{Chen04} can be used to address both these problems, that is via connection with entanglement distillation. To this end the wealth of results on entanglement distillation for graph states~\cite{HDERNB06} will help push this direction further.

We hope that this work will enable the integration of different quantum information protocols, for example secret computing.
perhaps the most interesting property of these schemes is their integratability.
As quantum technologies become more accessible in the real world, it is likely that in a real network setting we would like to do combinations of tasks, without having to change and swap hardware. The embedded protocols here are a first step towards this goal, and we hope these ideas can pave the way for more integrated schemes.

\acknowledgments
DM acknowledges financial support from QICS,
and BCS acknowledges financial support from CIFAR, MITACS, iCORE, and NSERC.
BCS appreciates valuable, seminal discussions with Anirban Roy and
Somshubhro Bandyopadhyay in the early stages of this work and
recent discussions with Vlad Gheorghiu and Robert B. Griffiths, who are engaged in
related research.
DM acknowledges enlightening discussions with Francesco Buscemi,
Peter van Loock, Mehdi Mhalla, and Elham Kashefi. We are very grateful to S.\ Perdrix, B.\ Fortescue and G.\ Gour for discovering our error in previous versions and pointing it out to us and to S.\ Perdrix for pointing out reference~\cite{BCT09};
the erratum for our original work has appeared~\cite{MS11}.

\end{document}